\documentclass[twocolumn]{revtex4}
\usepackage{amssymb}
\usepackage{latexsym}
\usepackage{epsfig}
\usepackage{subfigure}
\usepackage{makecell}
\usepackage{amsmath}
\usepackage[colorlinks=true,linkcolor=red]{hyperref}
\newcommand{\bea}{\begin{eqnarray}}
\newcommand{\eea}{\end{eqnarray}}
\newcommand{\beq}{\begin{equation}}
\newcommand{\eeq}{\end{equation}}

\begin{document}

\title{Dynamical analysis and statefinder of Barrow holographic dark energy}
\author{Qihong Huang$^{1}$\footnote{Corresponding author: huangqihongzynu@163.com}, He Huang$^2$, Bing Xu$^3$, Feiquan Tu$^{1}$ and Jun Chen$^{4}$ }
\affiliation{
$^1$ School of Physics and Electronic Science, Zunyi Normal University, Zunyi 563006, China\\
$^2$College of Physics and Electronic Engineering, Nanning Normal university, Nanning 530001, China\\
$^3$School of Electrical and Electronic Engineering, Anhui Science and Technology University, Bengbu 233030, China\\
$^4$ School of Science, Kaili University, Kaili, Guizhou 556011, China
}

\begin{abstract}
Based on the holographic principle and the Barrow entropy, Barrow holographic dark energy had been proposed. In order to analyze the stability and the evolution of Barrow holographic dark energy, we, in this paper, apply the dynamical analysis and statefinder methods to Barrow holographic dark energy with different IR cutoff and interacting terms. In the case of using Hubble horizon as IR cutoff with the interacting term $Q=\frac{\lambda}{H}\rho_{m}\rho_{D}$, we find this model is stable and can be used to describe the whole evolution of the universe when the energy transfers from the pressureless matter to the Barrow holographic dark energy. When the dynamical analysis method is applied to this stable model, an attractor corresponding to an accelerated expansion epoch exists and this attractor can behave as the cosmological constant. Furthermore, the coincidence problem can be solved in this case. Then, after using the statefinder analysis method to this model, we find this model can be discriminated from the standard $\Lambda$CDM model. Finally, we have discussed the turning point of Hubble diagram in Barrow holographic dark energy and find the turning point does not exist in this model.

\end{abstract}

\maketitle

\section{Introduction}

Observations~\cite{Perlmutter1999, Riess1998, Spergel2003, Spergel2007, Tegmark2004, Eisenstein2005} indicate that the universe is currently undergoing a phase of accelerated expansion. Nevertheless, Einstein's General Relativity with only radiation and matter cannot result in an accelerated expansion of the universe. In order to explain this observational result, a mysterious matter called dark energy was introduced to explain this accelerated phase. Since dark energy has not been detected directly, there is no reason to assume dark energy resembles known forms of matter or energy, and it becomes one of the biggest mysteries in modern cosmology. The simplest dark energy model is the cosmological constant ($\Lambda$CDM) in which the constant vacuum energy density can drive the accelerated expansion. Although it is well in agreement with the current observations~\cite{Planck2020}, it suffers from the fine-tuning problem and the coincidence problem.

An interesting approach to describe the origin of dark energy is considering the holographic principle~\cite{Susskind1995,Bousso2002} in the cosmological framework. After the holographic principle was introduced in cosmology, the holographic dark energy (HDE) arose~\cite{Cohen1999}. However, the original HDE is not successful since it cannot explain the current accelerated expansion of the universe~\cite{Hsu2004,Li2004}. To remedy this situation, a new HDE was proposed~\cite{Li2004}. Then, the HDE model becomes a physically viable dark energy candidate and has been widely studied theoretically~\cite{Horvat2004, Huang2004, Huang2005, Pavon2005, Wang2005, Zhang2006, Nojiri2006, Kim2006, Wang2006, Setare2006, Zhang2007, Wei2007, Li2008, Setare2009, Li2010, Gong2010, Zhang2010, Wang2017} and observationally~\cite{Zhang2005, Feng2007, Li2009, Zhang2009, Lu2010, Micheletti2010, Li2013, Agostino2019}. In HDE, various observational data strongly constrains the free parameter $c$ of HDE being less than 1~\cite{Huang2004a, Zhang2005, Shen2005, Chang2006, Zhang2007a, Yi2007, Ma2009, Li2009, Li2013, Feng2016, Akhlaghi2018, Da2020}, indicating HDE would lead to a phantom universe with big rip~\cite{Li2009a}. To solve the big rip, an interaction between dark energy and dark matter within HDE model was introduced~\cite{Li2009a, Zhang2012}. In HDE, this interaction was also used to alleviate the cosmic coincidence problem~\cite{Pavon2005, Duran2010, Oliveros2015} and resolved the classical instability~\cite{Oliveros2015} which stems from a negative sound speed~\cite{Myung2007}. And the interaction terms within HDE have been tested and constrained by various astronomical observations~\cite{Wang2006, Feng2007, Wu2008, Duran2010, Mukherjee2019, Kim2020} including Planck 2015~\cite{Feng2016}.

Horizon entropy is the most important cornerstone of HDE models, any change of it will lead to different HDE models. Inspired by the holographic principle and using different horizon entropy, some new HDE models were proposed recently, such as, Tsallis holographic dark energy (THDE)~\cite{Tavayef2018}, Renyi holographic dark energy~\cite{Moradpour2018}, Ricci-Gauss-Bonnet holographic dark energy~\cite{Saridakis2018} and Sharma-Mittal holographic dark energy~\cite{Jahromi2018}. Another cornerstone of HDE models is IR cutoff, different IR cutoffs can also result in new HDE models~\cite{Cohen1999, Li2004, Zadeh2018}. These new proposed HDE models were studied in various scenario~\cite{Zadeh2018, Saridakisa2018, Ghaffari2018, Huang2019, Sadri2019, Aditya2019, Mamon2020, Saha2020, Ebrahimi2020, Ens2020, Yadav2021, Dubey2020, Saha2021, Ahmed2020, Maity2019, Iqbal2019}.

Based on the generic holographic principle and using the Barrow entropy~\cite{Barrow2020} instead of the Bekenstein-Hawking entropy, Barrow holographic dark energy (BHDE) was proposed by considering the future event horizon as IR cutoff~\cite{Saridakis2020, Anagnostopoulos2020}. When the IR cutoff was chosen as the Hubble horizon, a new BHDE was proposed~\cite{Srivastava2021}. Then, the thermodynamics of this new BHDE with an interaction term was studied~\cite{Mamon2021}. However, this new BHDE is unstable since its squared sound speed is negative~\cite{Srivastava2021}. It is noted that an interaction term between dark energy and dark matter can be used to solve the classical instability in HDE~\cite{Oliveros2015} and THDE~\cite{Huang2019}, and this interaction term can also alleviate the cosmic coincidence problem~\cite{Pavon2005, Duran2010, Oliveros2015}. So, in BHDE models, it is unclear whether the classical instability and the cosmic coincidence problem can be solved when an interaction term is taken into account.

When a dark energy model is proposed, it should fit the conventional standard cosmology as well as explain the current accelerated expansion. For a viable cosmological model, it should describe the whole evolution history of the universe. Namely, the universe stems from a radiation dominated epoch, and then enters into a matter dominated epoch to enable the formation of large-scale structures, and eventually evolves into a dark energy dominated epoch. To analyze the evolution of the universe, the dynamical analysis method is introduced, which is an excellent method to investigate the qualitative behavior of a given cosmological model and can avoid the difficulty in solving non-linear cosmological equations. In this method, the dynamics of the universe can be described by analyzing the behavior of critical points of the dynamical system of a model, and the critical points always indicate the main evolution epoch of the universe. The stable point is used to describe a late time epoch dominated by the dark energy, the saddle point can denote the matter dominated epoch, and the unstable point corresponds to the early radiation dominated epoch. This method was used with great success in analyzing the evolution of the universe within HDE models~\cite{Setare2009a, Mahata2015, Caldera-Cabral2009, Banerjee2015, Liu2010, Ebrahimi2020, Huang2019}. These positive results in HDE lead to some interesting questions in BHDE and motivate us to study BHDE: Whether a stable BHDE model can describe the whole evolution of the universe. How to discriminate the BHDE model from the standard $\Lambda$CDM model if the BHDE model describes the cosmic evolution successfully? The main task of this paper is to answer these questions.

Furthermore, a turning point of Hubble diagram in HDE was studied in Ref.~\cite{Colgain2021} which shows that HDE model may be at odds with the cosmological principle. Here, the free parameter region $0<c<1$, which was constrained by the current data, was considered. For $0.5<c<1$, this turning point exists in the future, while it occurs in the past and is observable for $c<0.5$. In BHDE with the future event horizon as IR cutoff, the combination $H(z)+SNIa$ place constraints on the free parameter $C=3.421^{+1.753}_{-1.611}$ and the parameter of Barrow entropy $\Delta=0.094^{+0.094}_{-0.101}$~\cite{Anagnostopoulos2020}. It is interesting to discuss whether this turning point exists in BHDE with the future event horizon as IR cutoff when the constraints $C=3.421^{+1.753}_{-1.611}$ and $\Delta=0.094^{+0.094}_{-0.101}$ are considered.

Regarding that BHDE may offer an interesting framework to study phenomenology beyond to HDE, the main goal of this paper is to discuss six BHDE models: we first consider BHDE with different IR cutoffs, and then, two kinds of interaction terms are used. The analysis is performed in four respects: Firstly, by analyzing the squared sound speed, we study the stability of BHDE models. Secondly, we use the dynamical analysis method to analyze the phase space behavior of the stable BHDE models and discuss the cosmic coincidence problem. Thirdly, in order to discriminate the stable BHDE model from the standard $\Lambda$CDM model, we perform the statefinder diagnostic by depicting the evolution trajectories of statefinder parameters. Finally, we plot the evolution curves of $H(z)$ and study the turning point of Hubble diagram in BHDE models.

The paper is organized as follows. In Section II, we investigate the evolution and stability of the universe in BHDE with different IR cutoff and interacting terms. In Section III, we use the dynamical analysis method to analyze the phase space behavior of the stable BHDE model. In Section IV, the statefinder diagnostic pairs are used to discriminate the stable BHDE model from the standard $\Lambda$CDM model. In Section V, we discuss the turning point of Hubble diagram in BHDE. Finally, our main conclusions are presented in Section VI.

\section{The Universe evolution}

Using the Barrow entropy~\cite{Barrow2020}, the energy density of BHDE is given as~\cite{Saridakis2020}
\beq
\rho_{D}=C L^{\triangle-2}.
\eeq
Here, $C$ is a parameter with the dimension $[L]^{-2-\triangle}$ and $\Delta$ satisfies the relation $0 \leq \Delta \leq 1$. For the case $\Delta=0$, $\rho_{D}$ provides a smooth spacetime structure, i.e. the standard holographic dark energy, and $\Delta=1$ corresponding to the most intricate structure.

We consider a homogeneous and isotropic flat Friedmann-Robertson-Walker universe
\beq
ds^{2}=-dt^{2}+a^{2}(t)(dr^{2}+r^{2}d\Omega^{2}),
\eeq
where $a(t)$ is the scale factor with $t$ being cosmic time, and setting $\kappa^{2}=8\pi G$. The Friedmann equation is given as
\beq\label{H2}
H^{2}=\frac{\kappa^{2}}{3}(\rho_{r}+\rho_{m}+\rho_{D}),
\eeq
where $\rho_{r}$, $\rho_{m}$ and $\rho_{D}$ denote the energy density of radiation, pressureless matter and BHDE, respectively. The radiation, pressureless matter and BHDE conservation equations take the form
\bea
&&\dot{\rho}_{r}+4H \rho_{r}=0,\label{r}\\
&&\dot{\rho}_{m}+3H \rho_{m}=Q,\label{m}\\
&&\dot{\rho}_{D}+3H(1+w_{D})\rho_{D}=-Q,\label{D}
\eea
in which $w_{D}=\frac{p_{D}}{\rho_{D}}$ denotes the equation of state parameter of BHDE and $Q$ represents the energy exchange between pressureless matter and BHDE. For $Q>0$, energy transfers from BHDE to pressureless matter, and energy transfers from pressureless matter to BHDE for $Q<0$. In this paper, we consider two interacting cases $Q=H(\alpha\rho_{m}+\beta\rho_{D})$~\cite{Caldera-Cabral2009, Huang2019, Bahamonde2018} and $Q=\frac{\lambda}{H}\rho_{m}\rho_{D}$~\cite{Perez2014, Bahamonde2018}. Here, $\alpha$, $\beta$ and $\lambda$ are coupling constants.

After introducing the following dimensionless density parameters
\bea\label{Orm}
&&\Omega_{r}=\frac{\kappa^{2}\rho_{r}}{3H^{2}}, \qquad \Omega_{m}=\frac{\kappa^{2}\rho_{m}}{3H^{2}}, \qquad \Omega_{D}=\frac{\kappa^{2}\rho_{D}}{3H^{2}},\nonumber \\
&&\sigma=\frac{\kappa^{2}Q}{3H^{3}},
\eea
we can express the Friedmann equation~(\ref{H2}) as
\beq\label{Orm1}
\Omega_{r}+\Omega_{m}+\Omega_{D}=1.
\eeq

Differentiating the Friedmann equation~(\ref{H2}) and combining Eqs.~(\ref{r}), ~(\ref{m}), ~(\ref{D}), ~(\ref{Orm}) and ~(\ref{Orm1}), we obtain
\beq\label{HH2}
\frac{\dot{H}}{H^{2}}=\frac{1}{2}[\Omega_{m}+(1-3w_{D})\Omega_{D}]-2.
\eeq
And the deceleration parameter $q$ can be calculated as
\beq
q=-1-\frac{\dot{H}}{H^{2}}=1-\frac{1}{2}[\Omega_{m}+(1-3w_{D})\Omega_{D}].
\eeq

Defining $\Omega'=d\Omega/d(lna)$ and differentiating Eq.~(\ref{Orm}), we get
\bea
&&\Omega^{'}_{m}=[(3w_{D}-1)\Omega_{D}-\Omega_{m}+1]\Omega_{m}+\sigma,\label{mD1}\\
&&\Omega^{'}_{D}=[(3w_{D}-1)(\Omega_{D}-1)-\Omega_{m}]\Omega_{D}-\sigma.\label{mD2}
\eea
Here, Eqs.~(\ref{m}), ~(\ref{D}), ~(\ref{Orm}), ~(\ref{Orm1}) and ~(\ref{HH2}) are used.

To discuss the stability of BHDE model, we need to analyze its squared sound speed
\beq
v^{2}_{s}=\frac{dp_{D}}{d\rho_{D}}=\frac{\dot{p}_{D}}{\dot{\rho}_{D}}=\frac{\rho_{D}}{\dot{\rho}_{D}}\dot{\omega}_D+\omega_{D}.
\eeq
For $v^{2}_{s}>0$, this model can be stable against perturbations. Otherwise, it is unstable.

To find a stable and suitable model for BHDE, we will analyze six models in the following:

(i)Model I: Non-interacting BHDE with future event horizon as IR cutoff (BHDEF);

(ii)Model II: Interacting BHDE with future event horizon as IR cutoff and $Q=H(\alpha\rho_{m}+\beta\rho_{D})$ (IBHDEFA);

(iii)Model III: Interacting BHDE with future event horizon as IR cutoff and $Q=\frac{\lambda}{H}\rho_{m}\rho_{D}$ (IBHDEFL);

(iv)Model IV: Non-interacting BHDE with Hubble horizon as IR cutoff (BHDEH);

(v)Model V: Interacting BHDE with Hubble horizon as IR cutoff and $Q=H(\alpha\rho_{m}+\beta\rho_{D})$ (IBHDEHA);

(vi)Model VI: Interacting BHDE with Hubble horizon as IR cutoff and $Q=\frac{\lambda}{H}\rho_{m}\rho_{D}$ (IBHDEHL).

\subsection{BHDE with future event horizon}

Considering the future event horizon as IR cutoff, the energy density of BHDE takes the form~\cite{Saridakis2020}
\beq\label{rhoD}
\rho_{D}=C R^{\triangle-2}_{h},
\eeq
with
\beq
R_{h}=a\int^{\infty}_{t}\frac{dt}{a}=a\int^{\infty}_{a}\frac{da}{H a^{2}},
\eeq
which satisfies the relation $\dot{R}_{h}=H R_{h}-1$. For the case $\Delta=0$ with $C=3c^{2}$, it reduces to the standard holographic dark energy.

In the following, we assume the present scale factor $a_{0}=1$ so that the redshift $z$ satisfies the relation $z=\frac{1}{a}-1$. Since the combination $H(z)+SNIa$ constrain $C=3.421^{+1.753}_{-1.611}$ and $\Delta=0.094^{+0.094}_{-0.101}$~\cite{Anagnostopoulos2020}, we consider $C=3.5$ and $\Delta=0.1$ within BHDE with future event horizon as IR cutoff.

\subsubsection{Non-interacting}

When there is no energy exchange between the pressureless matter and BHDE, we obtain $Q=0$ and $\sigma=0$. By differentiating Eq.~(\ref{rhoD}) and using Eq.~(\ref{D}), one can obtain the equation of state parameter of BHDE
\beq\label{wD1}
w_{D}=\frac{\Delta-2}{3}F-\frac{\Delta+1}{3},
\eeq
with $F=\frac{1}{H R_{h}}$ and $F'$ satisfies
\beq\label{F1}
F^{'}=-\Big[\frac{1}{2}(\Omega_{m}+(1-3w_{D})\Omega_{D})-F-1\Big]F.
\eeq
Using Eqs.~(\ref{rhoD}) and~(\ref{wD1}), the squared sound speed can be written as
\beq
v^{2}_{s}=\frac{F'}{3(1-F)}+\frac{\Delta-2}{3}F-\frac{\Delta+1}{3}.
\eeq
Substituting Eq.~(\ref{F1}) into the above relation, $v^{2}_{s}>0$ leads to
\begin{small}
\beq
0 \leq \Omega_{D} < \frac{4-\Omega_{m}}{4}, \frac{\Omega_{m}+(\Delta+2)\Omega_{D}-4\Delta}{2[(\Delta-2)\Omega_{D}+2(3-\Delta)]}+\frac{1}{2}\sqrt{\xi}<F<1,\nonumber
\eeq
\end{small}
or
\begin{small}
\beq
\frac{4-\Omega_{m}}{4} < \Omega_{D} \leq 1, 1<F<\frac{\Omega_{m}+(\Delta+2)\Omega_{D}-4\Delta}{2[(\Delta-2)\Omega_{D}+2(3-\Delta)]}+\frac{1}{2}\sqrt{\xi},\nonumber
\eeq
\end{small}
in which
\bea
&&\xi=\frac{\Omega_{m}^{2}+[2(\Delta+2)\Omega_{D}-8\Delta]\Omega_{m}}{[(\Delta-2)\Omega_{D}+2(3-\Delta)]^{2}}\nonumber\\
&&\quad+\frac{(2+\Delta)^{2}\Omega_{D}^{2}-8(3\Delta+2)\Omega_{D}+16(3+2\Delta)}{[(\Delta-2)\Omega_{D}+2(3-\Delta)]^{2}}.\nonumber
\eea
These results mean the condition $v^{2}_{s}>0$ cannot be satisfied during the whole evolution of the universe. So, this model is unstable.

\begin{figure*}[htp]
\begin{center}
\includegraphics[width=0.45\textwidth]{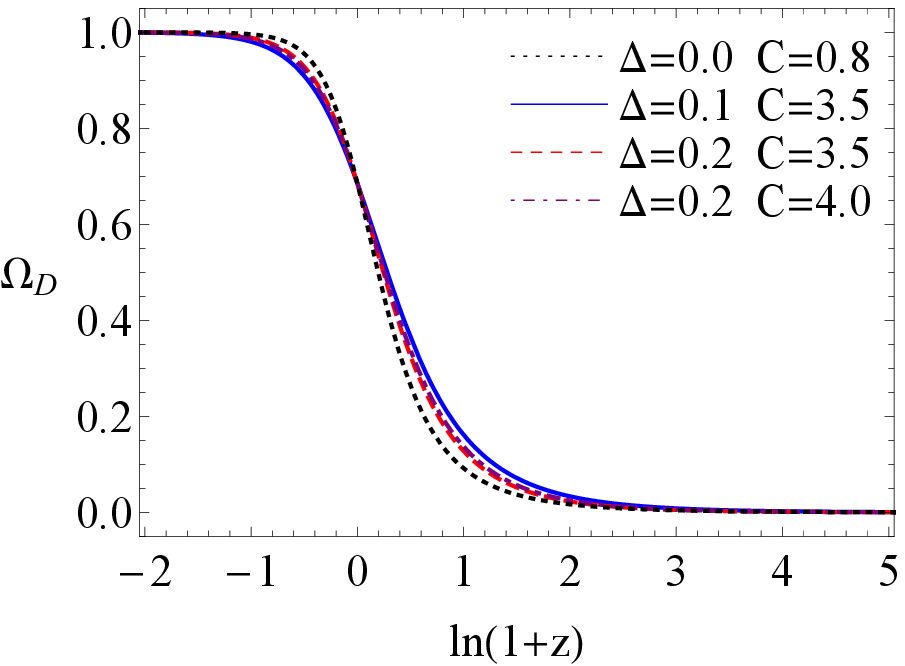}
\includegraphics[width=0.45\textwidth]{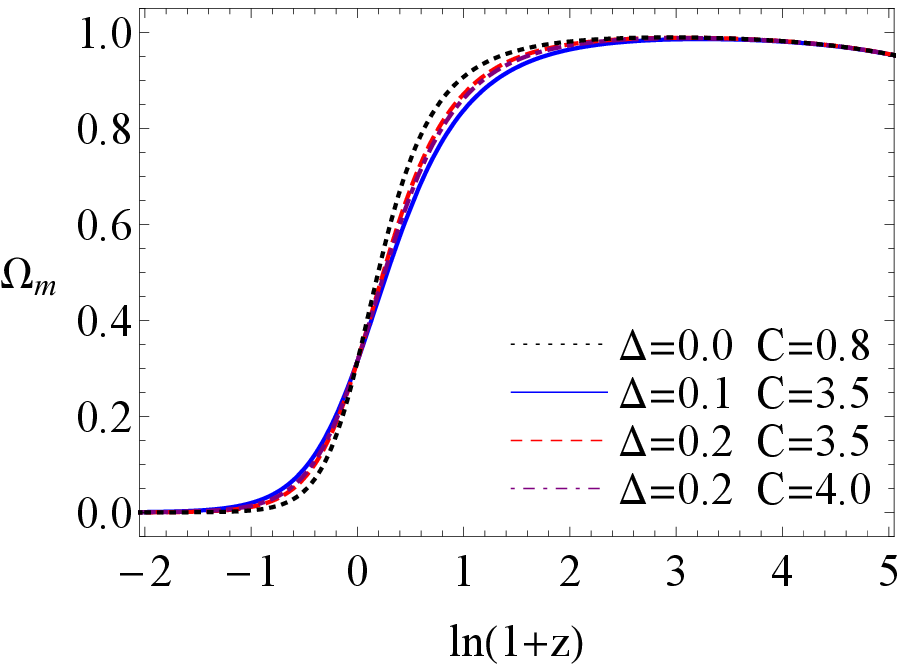}
\caption{\label{Fig01} Evolution curves of $\Omega_{D}$ and $\Omega_{m}$ versus redshift parameter $ln(1+z)$ for BHDEF.}
\end{center}
\end{figure*}

\begin{figure*}[htp]
\begin{center}
\includegraphics[width=0.45\textwidth]{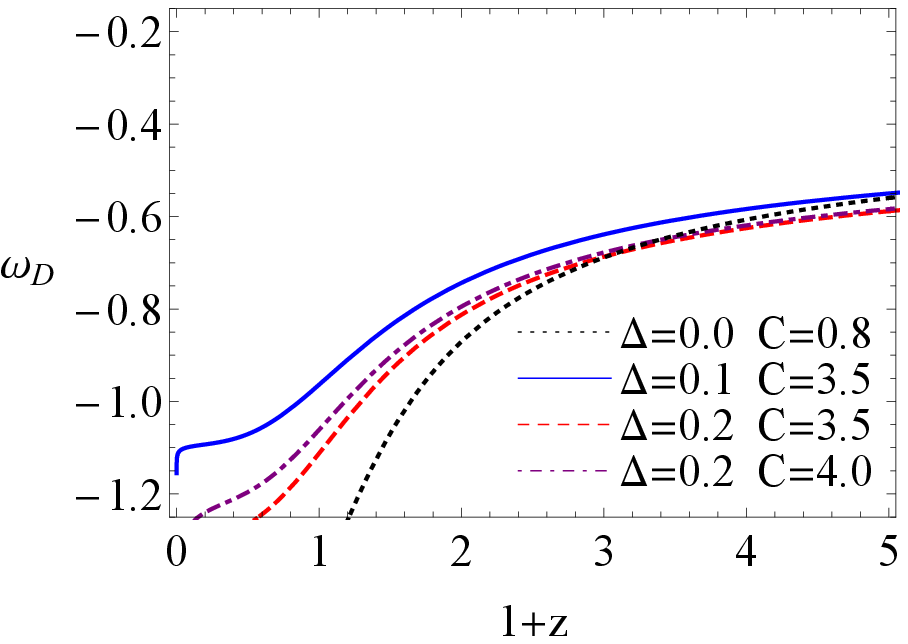}
\includegraphics[width=0.435\textwidth]{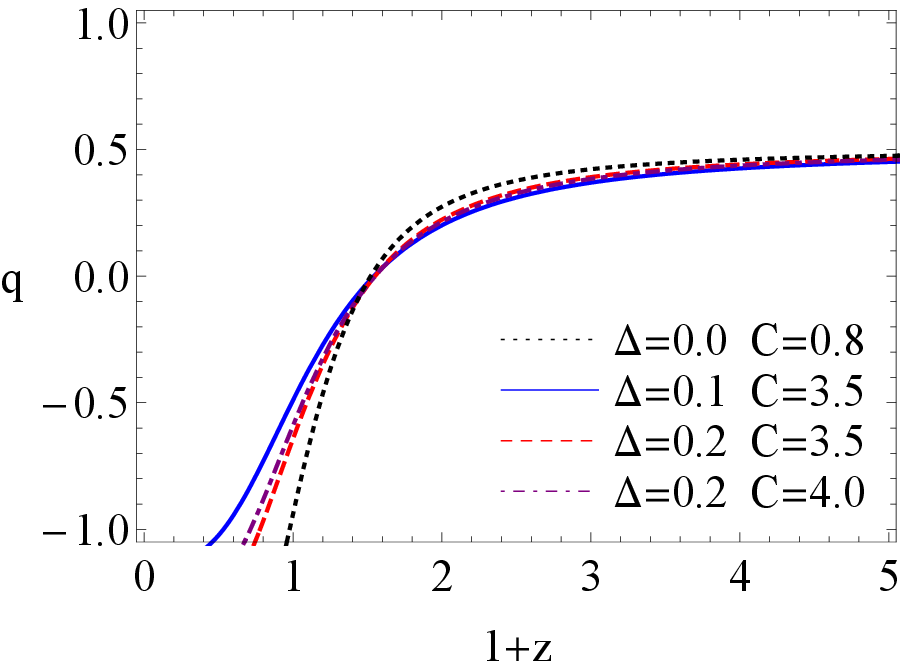}
\caption{\label{Fig02} Evolution curves of $w_{D}$ and $q$ versus redshift parameter $1+z$ for BHDEF.}
\end{center}
\end{figure*}

\begin{figure*}[htp]
\begin{center}
\includegraphics[width=0.45\textwidth]{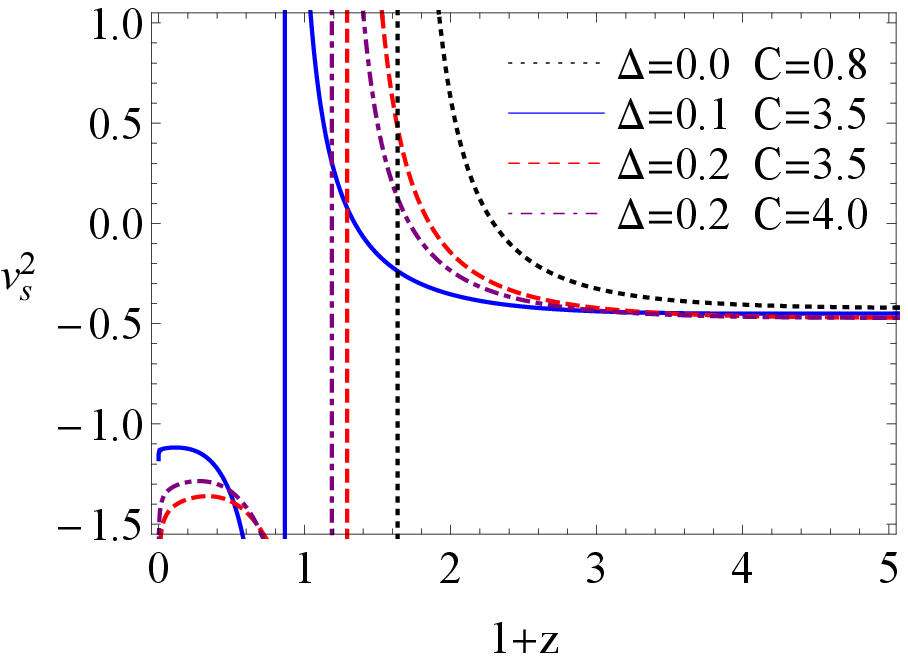}
\includegraphics[width=0.45\textwidth]{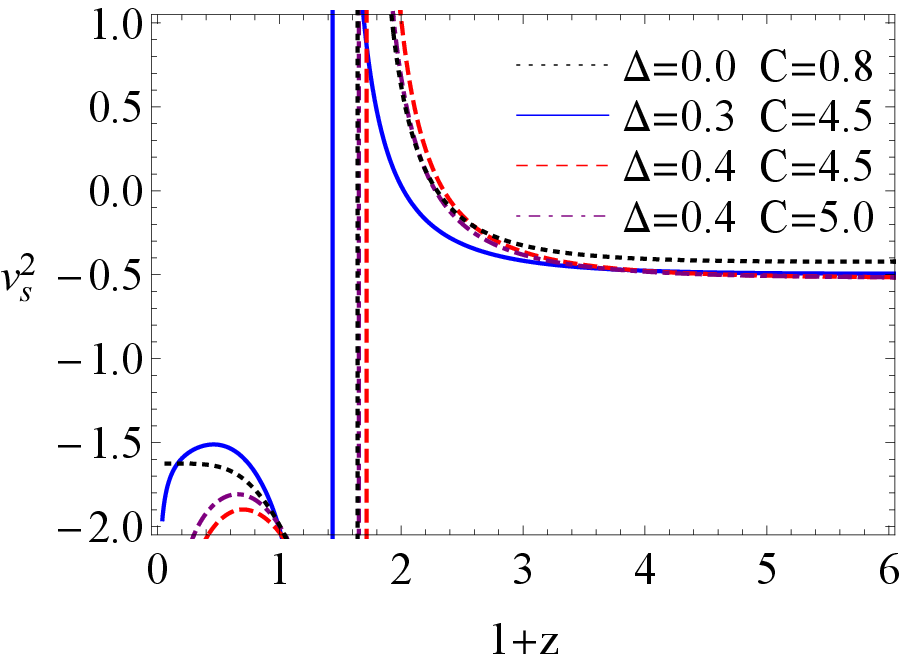}
\caption{\label{Fig03} Evolution curves of $v_{s}^{2}$ versus redshift parameter for BHDEF.}
\end{center}
\end{figure*}

Throughout this paper, we choose $\Omega^{0}_{r}=0.0001$, $\Omega^{0}_{m}=0.3152$, $\Omega^{0}_{D}=0.6847$ and $H_{0}=67.36 km s^{-1}Mpc^{-1}$~\cite{Planck2020} as the initial conditions. For the cases of $\Delta=0.1, C=3.5$, $\Delta=0.2, C=3.5$ and $\Delta=0.2, C=4.0$ with the initial conditions, by solving Eqs.~(\ref{mD1}), ~(\ref{mD2}) and~(\ref{F1}) numerically, we get the evolution curves of $\Omega_{D}$, $\Omega_{m}$, $w_{D}$ and $q$ which are plotted in Figs.~(\ref{Fig01}) and ~(\ref{Fig02}). And the black dotted line in Figs.~(\ref{Fig01}) and ~(\ref{Fig02}) represents the situation of $\Delta=0$ which corresponds to the standard entropy.

Form Fig.~(\ref{Fig01}), we can see $\Omega_{D}\rightarrow 1$ and $\Omega_{m}\rightarrow 0$ at the late time, and the universe can be dominated by BHDE at the late time evolution. The left panel of Fig.~(\ref{Fig02}) shows that the BHDE behave as the quintessence and $w_{D}$ can cross the phantom line. The right panel of Fig.~(\ref{Fig02}) indicates a suitable range for the transition redshift is obtainable. But this model is unstable since $v^{2}_{s}>0$ cannot be satisfied, and some evolution curves of $v_{s}^{2}$ are depicted in Fig.~(\ref{Fig03}).

\subsubsection{Interacting with $Q=H(\alpha\rho_{m}+\beta\rho_{D})$}

When the energy exchange $Q=H(\alpha\rho_{m}+\beta\rho_{D})$ between the pressureless matter and BHDE exists, the equation of state parameter $w_{D}$ has the form
\beq
w_{D}=\frac{\Delta-2}{3}F-\frac{\alpha\Omega_{m}+\beta\Omega_{D}}{3\Omega_{D}}-\frac{\Delta+1}{3}.
\eeq
And the squared sound speed $v^{2}_{s}$ is given as
\bea
&&v_{s}^{2}=\frac{(\Delta-2)\Omega_{D}^{2}F'-\alpha \Omega_{m}' \Omega_{D}+\alpha \Omega_{m} \Omega_{D}'}{3(\Delta-2)(1-F)\Omega_{D}^{2}}\nonumber\\
&&\quad+\frac{\Delta-2}{3}F-\frac{\alpha \Omega_{m}+\beta \Omega_{D}}{3\Omega_{D}}-\frac{1+\Delta}{3}.
\eea
Here, $F'$ is given in Eq.~(\ref{F1}). Since the expression of $v^{2}_{s}$ is too complicated, we cannot get the analytical solution for $v^{2}_{s}>0$ and then resort to numerical solution.

\begin{figure*}[htp]
\begin{center}
\includegraphics[width=0.45\textwidth]{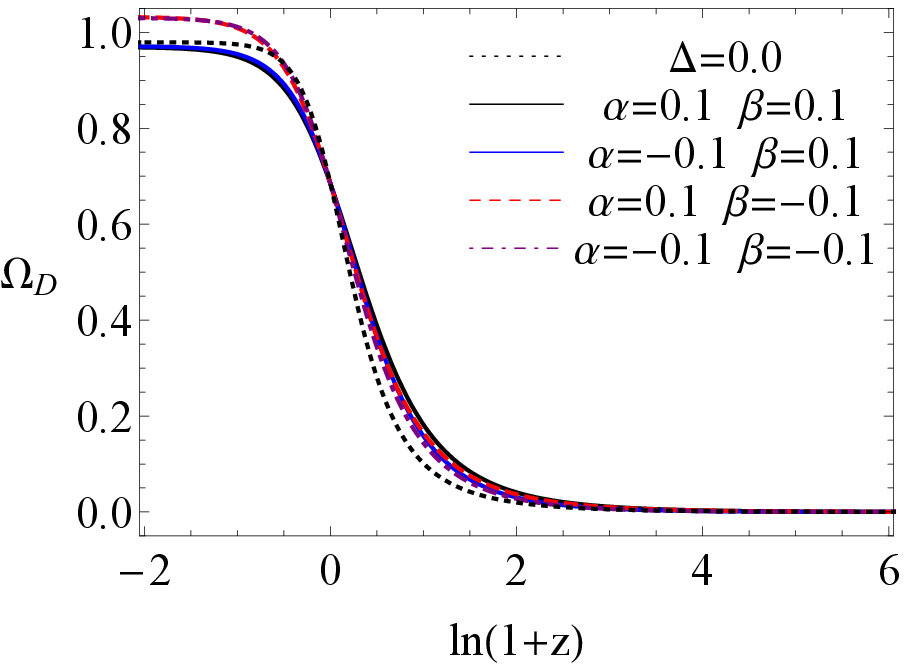}
\includegraphics[width=0.45\textwidth]{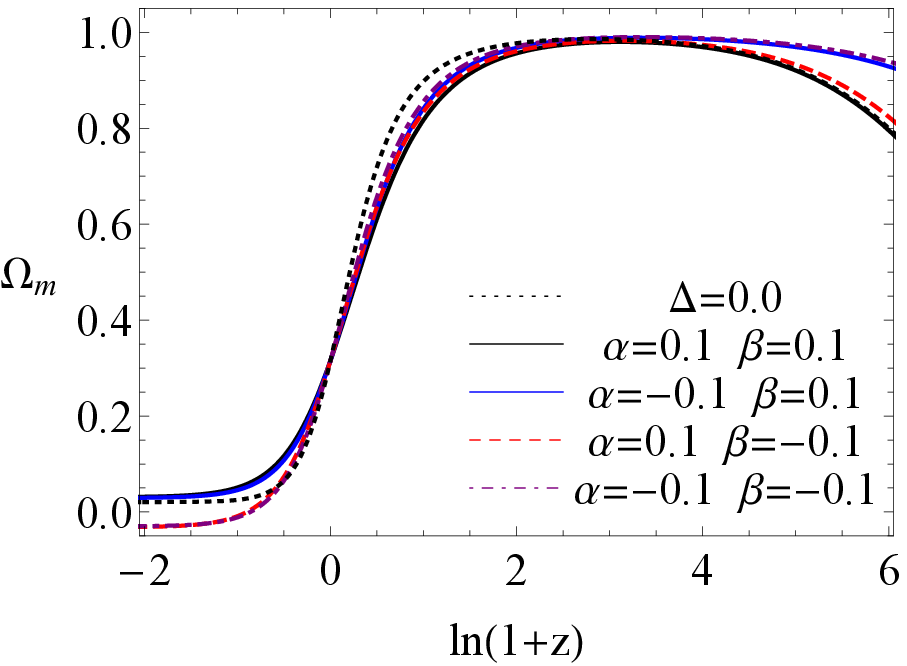}
\caption{\label{Fig04} Evolution curves of $\Omega_{D}$ and $\Omega_{m}$ versus redshift parameter $ln(1+z)$ for IBHDEFA.}
\end{center}
\end{figure*}

\begin{figure*}[htp]
\begin{center}
\includegraphics[width=0.45\textwidth]{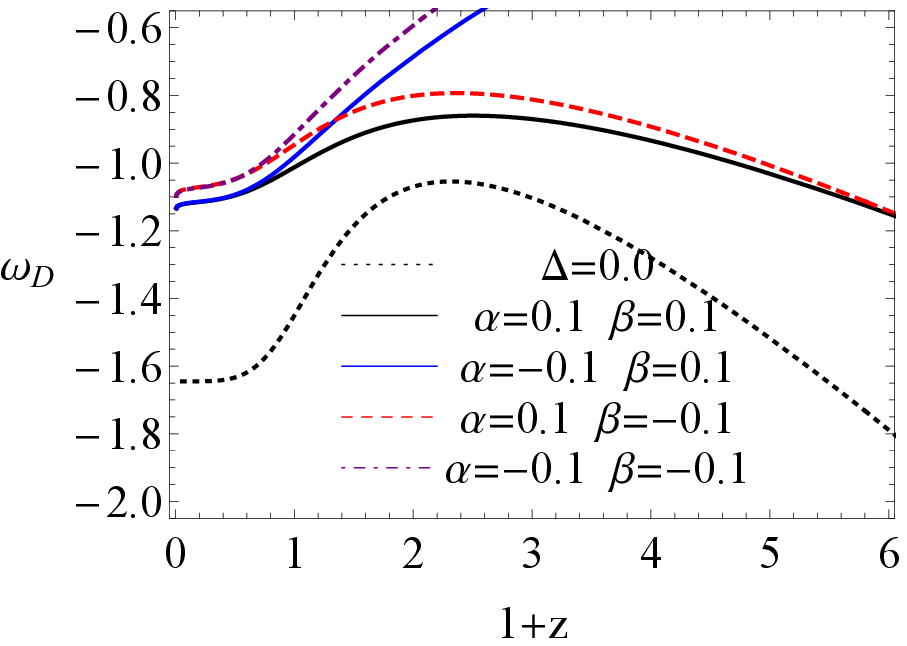}
\includegraphics[width=0.435\textwidth]{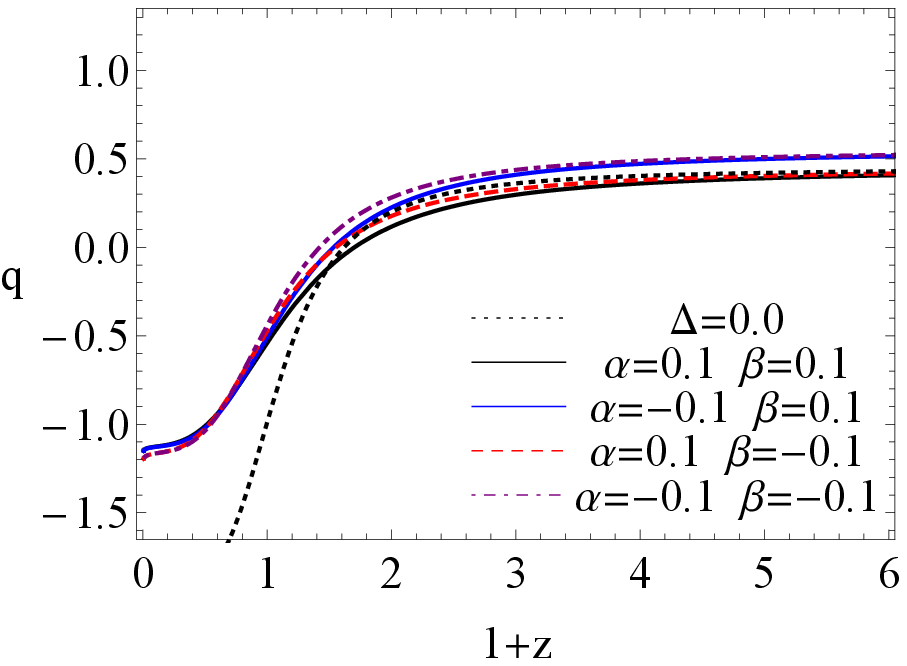}
\caption{\label{Fig05} Evolution curves of $w_{D}$ and $q$ versus redshift parameter $1+z$ for IBHDEFA.}
\end{center}
\end{figure*}

\begin{figure*}[htp]
\begin{center}
\includegraphics[width=0.45\textwidth]{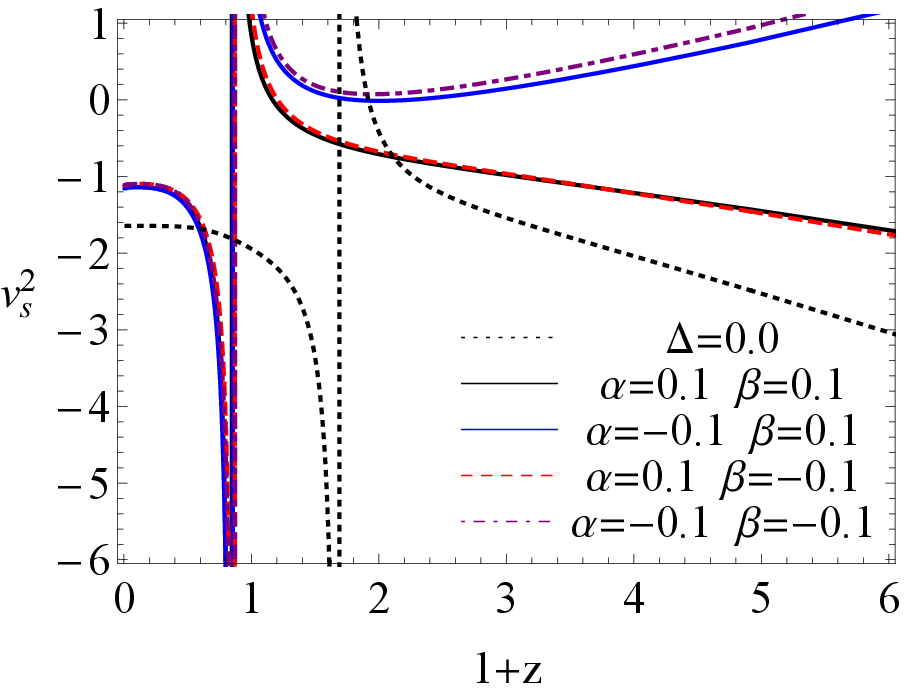}
\includegraphics[width=0.45\textwidth]{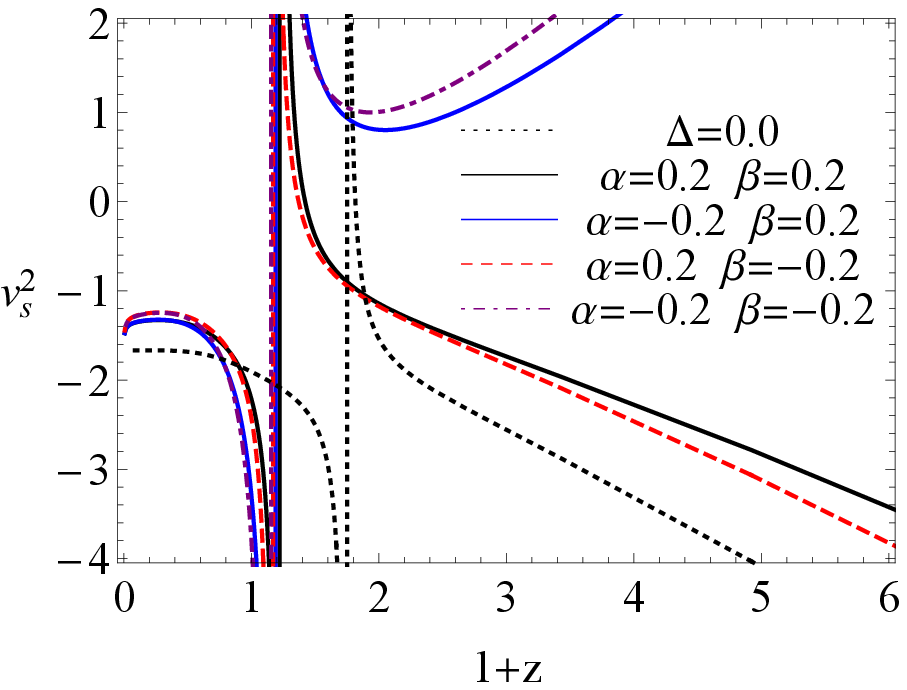}
\caption{\label{Fig06} Evolution curves of $v_{s}^{2}$ versus redshift parameter for IBHDEFA. The left panel was plotted for $\Delta=0.1,C=3.5$, while the right one was shown for $\Delta=0.2,C=4.0$.}
\end{center}
\end{figure*}

Using the initial conditions with $\Delta=0.1, C=3.5$ and solving Eqs.~(\ref{mD1}), ~(\ref{mD2}) and~(\ref{F1}), we obtain the evolution curves for $\Omega_{D}$, $\Omega_{m}$, $w_{D}$, $q$ and $v_{s}^{2}$ which are plotted in Figs.~(\ref{Fig04}), ~(\ref{Fig05}) and~(\ref{Fig06}). The specific case of $\Delta=0$ is plotted by the black dotted line, and the values of $\alpha$, $\beta$ and $C$ are $0.1$, $0.1$ and $0.8$, respectively.

It can be seen from Fig.~(\ref{Fig04}) where $\Omega_{D}$ is larger than $1$ and $\Omega_{m}$ is less than $0$ for a negative $\beta$. And for the positive $\beta$, the universe can be dominated by BHDE at the late time. The evolution curves of $w_{D}$ and $q$ is shown in Fig.~(\ref{Fig05}). The left panel of Fig.~(\ref{Fig05}) shows that BHDE can behave as quintessence and cross the phantom line for a negative $\alpha$, and the right one shows that the late time acceleration can be achieved and the transition redshift is obtainable. These results mean that $\alpha<0$ and $\beta>0$ can be used to describe the evolution of the universe. However, the result from Fig.~(\ref{Fig06}) means this model is unstable since the evolution curve of $v^{2}_{s}$ is not always positive during the whole evolution epoch.

\subsubsection{Interacting with $Q=\frac{\lambda}{H}\rho_{m}\rho_{D}$}

When the energy exchange $Q=\frac{\lambda}{H}\rho_{m}\rho_{D}$ between the pressureless matter and BHDE is adopted, $w_{D}$ and $v^{2}_{s}$ are written as
\beq
w_{D}=\frac{\Delta-2}{3}F-\lambda \Omega_{m}-\frac{\Delta+1}{3}.
\eeq
and
\bea
&&v^{2}_{s}=\frac{F'}{3(1-F)}+\frac{\Delta-2}{3}F-\frac{\Delta+1}{3}\nonumber\\
&&\qquad -\frac{\lambda\Omega_{m}'}{(\Delta-2)(1-F)}-\lambda\Omega_{m}.
\eea
Here, $F'$ is given by Eq.~(\ref{F1}).

\begin{figure*}[htp]
\begin{center}
\includegraphics[width=0.45\textwidth]{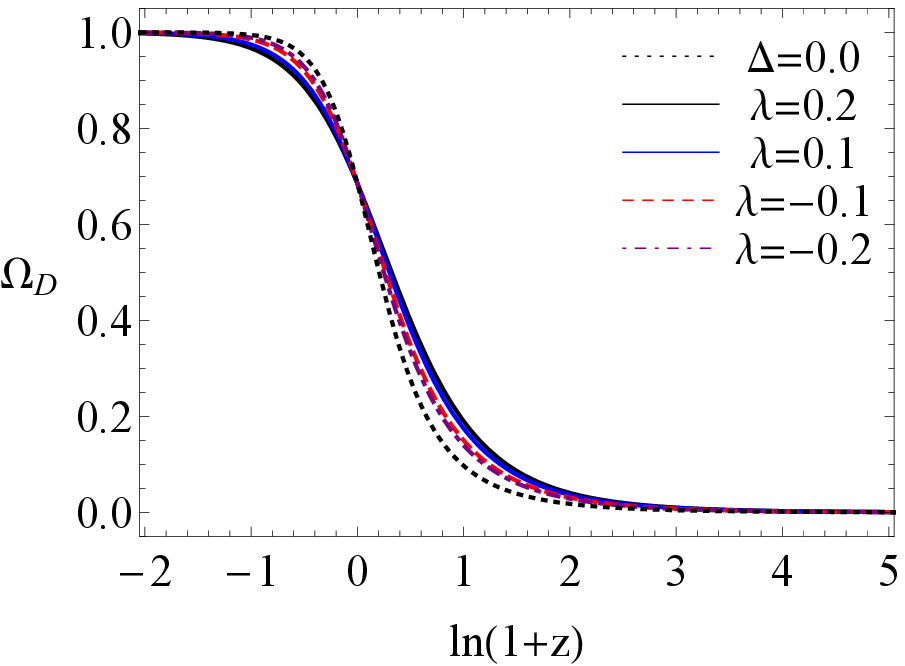}
\includegraphics[width=0.45\textwidth]{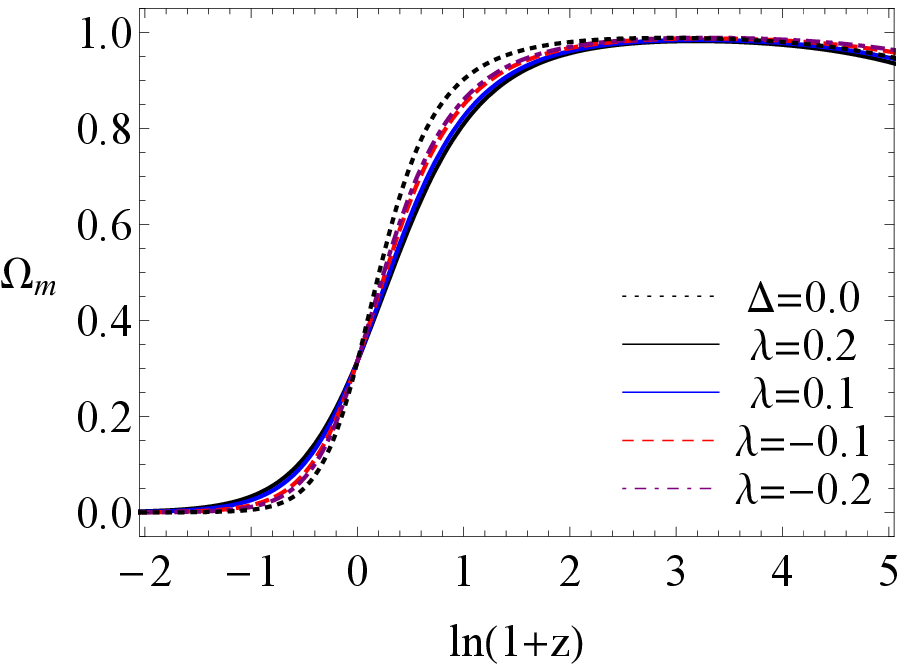}
\caption{\label{Fig07} Evolution curves of $\Omega_{D}$ and $\Omega_{m}$ versus redshift parameter $ln(1+z)$ for IBHDEFL.}
\end{center}
\end{figure*}

\begin{figure*}[htp]
\begin{center}
\includegraphics[width=0.45\textwidth]{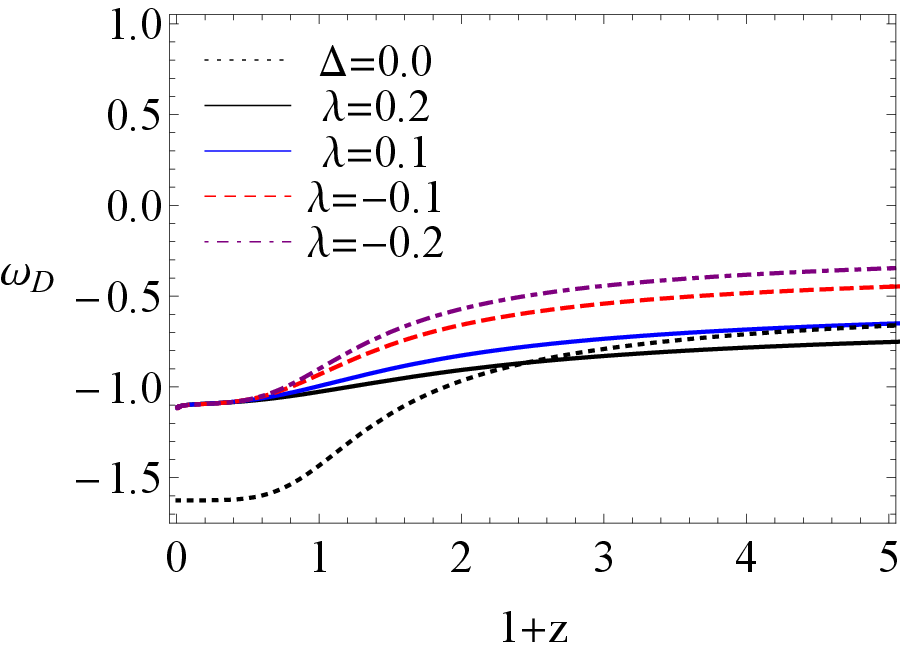}
\includegraphics[width=0.435\textwidth]{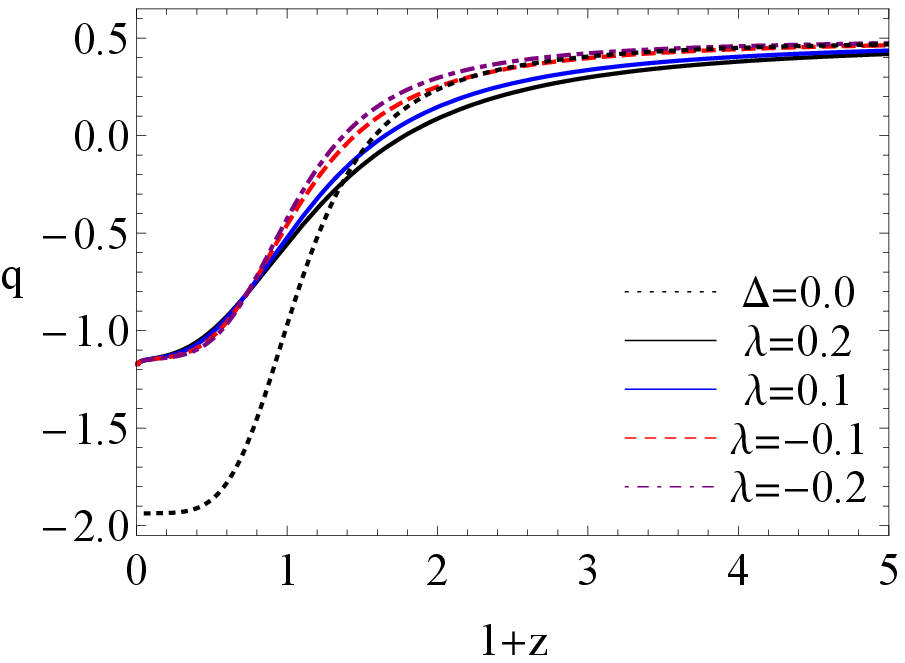}
\caption{\label{Fig08} Evolution curves of $w_{D}$ and $q$ versus redshift parameter $1+z$ for IBHDEFL.}
\end{center}
\end{figure*}

\begin{figure*}[htp]
\begin{center}
\includegraphics[width=0.45\textwidth]{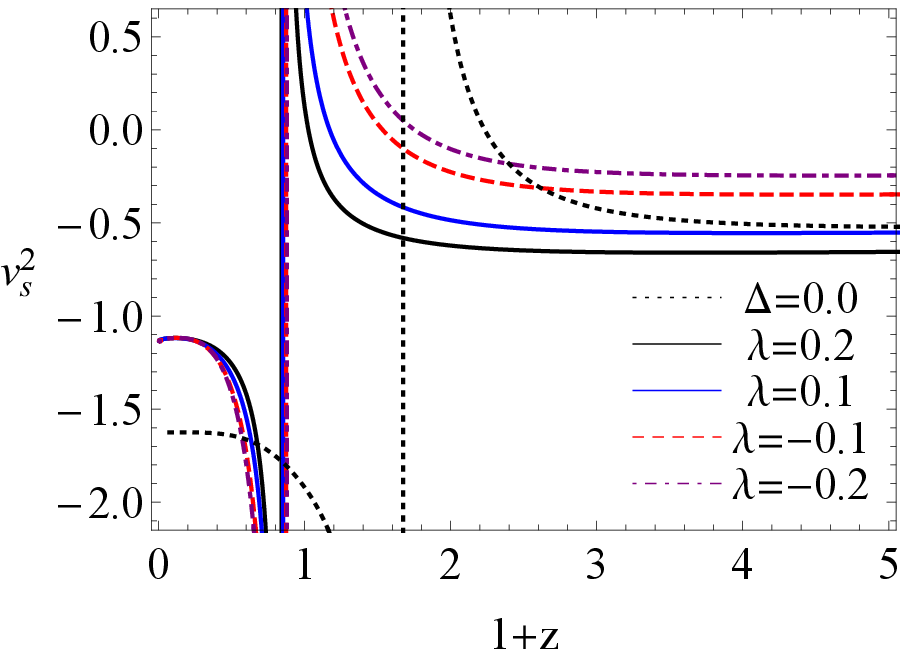}
\includegraphics[width=0.45\textwidth]{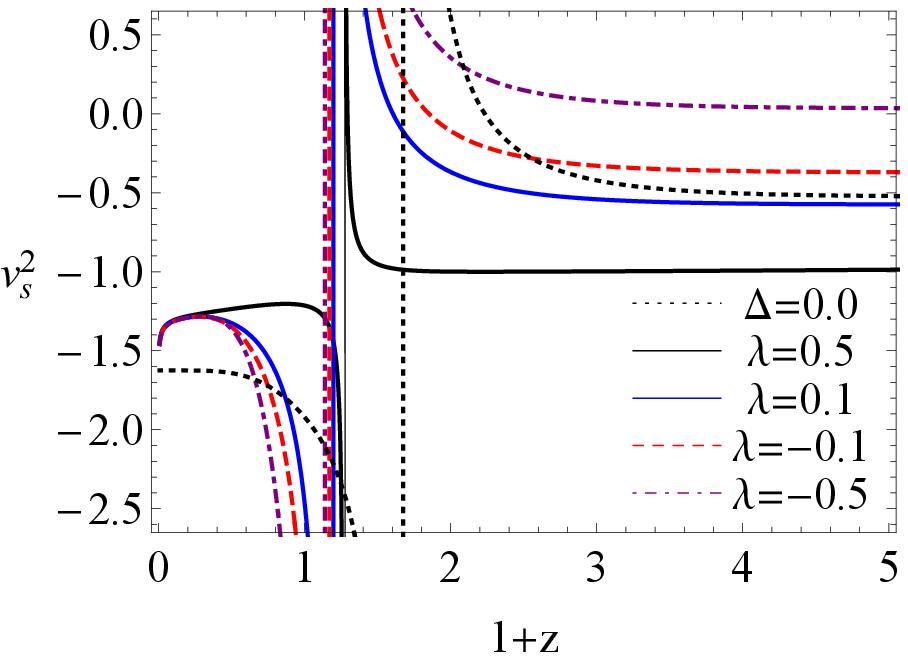}
\caption{\label{Fig09} Evolution curves of $v_{s}^{2}$ versus redshift parameter for IBHDEFL. The left panel was plotted for $\Delta=0.1,C=3.5$, while the right one was shown for $\Delta=0.2,C=4.0$.}
\end{center}
\end{figure*}

Then, solving Eqs.~(\ref{mD1}), ~(\ref{mD2}) and~(\ref{F1}) with $\Delta=0.1$ and $C=3.5$, we get the evolution curves of $\Omega_{D}$, $\Omega_{m}$, $w_{D}$, $q$ and $v_{s}^{2}$ which are plotted in Figs.~(\ref{Fig07}), ~(\ref{Fig08}) and~(\ref{Fig09}). The black dotted line in Figs.~(\ref{Fig07}), ~(\ref{Fig08}) and~(\ref{Fig09}) denotes the specific case of $\Delta=0$ with $\lambda=0.1$ and $C=0.8$. From these figures, one can see that this model can produce suitable results if proper values are chosen, but it is unstable.

\subsection{BHDE with Hubble horizon}

Using the Hubble horizon as IR cutoff, the energy density of BHDE leads to
\beq\label{rhoDH}
\rho_{D}=C H^{2-\triangle}.
\eeq
For $\Delta=2(\delta-1)$, the energy density of BHDE has the same form as that of Tsallis holographic dark energy (THDE) in mathematics~\cite{Tavayef2018}. Different from THDE, the parameter $\Delta$ in BHDE is limited to $0\leq \Delta \leq 1$.

\subsubsection{Non-interacting}

For the case $Q=0$, differentiating Eq.~(\ref{rhoDH}) and using Eq.~(\ref{D}), the equation of state parameter $w_{D}$ and the squared sound speed $v^{2}_{s}$ lead to
\beq
w_{D}=\frac{(\Delta-2)(\Omega_{m}+\Omega_{D}-1)-3\Delta}{3(\Delta-2)\Omega_{D}+6}.
\eeq
and
\beq
v^{2}_{s}=\frac{4(\Omega_{m}+\Omega_{D}-1)}{3(\Omega_{m}+4\Omega_{D}-4)}+\frac{2\Delta(\Omega_{m}+4\Omega_{D}-4)}{3[(\Delta-2)\Omega_{D}+2]^{2}}.
\eeq

\begin{figure*}[htp]
\begin{center}
\includegraphics[width=0.45\textwidth]{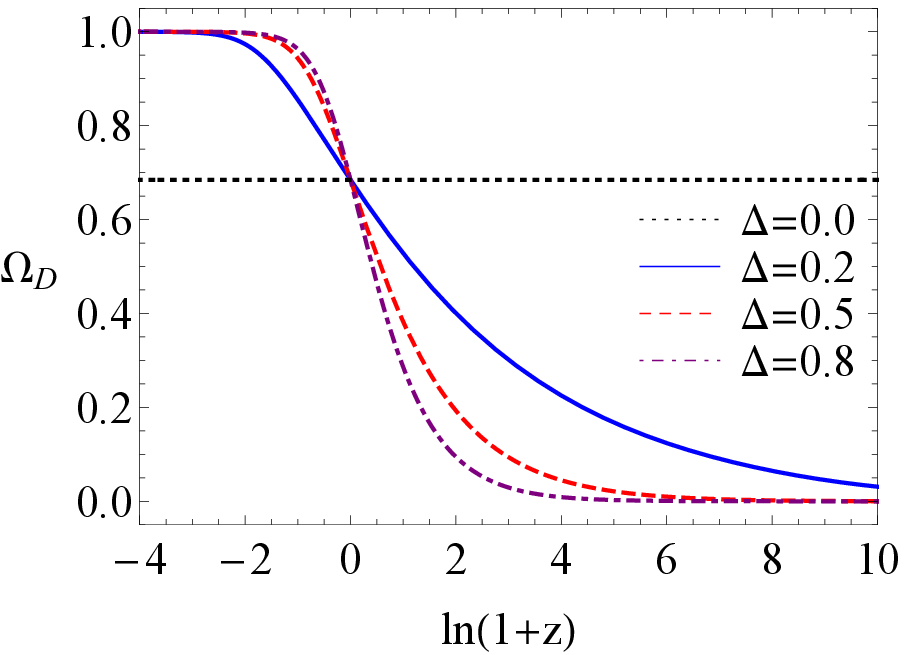}
\includegraphics[width=0.45\textwidth]{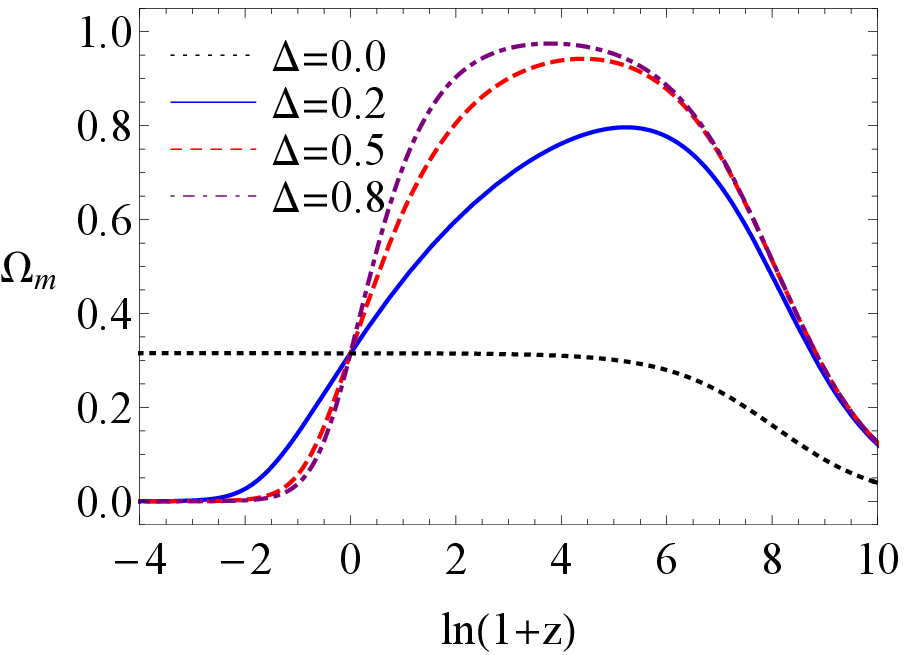}
\caption{\label{Fig10} Evolution curves of $\Omega_{D}$ and $\Omega_{m}$ versus redshift parameter $ln(1+z)$ for BHDEH.}
\end{center}
\end{figure*}

\begin{figure*}[htp]
\begin{center}
\includegraphics[width=0.45\textwidth]{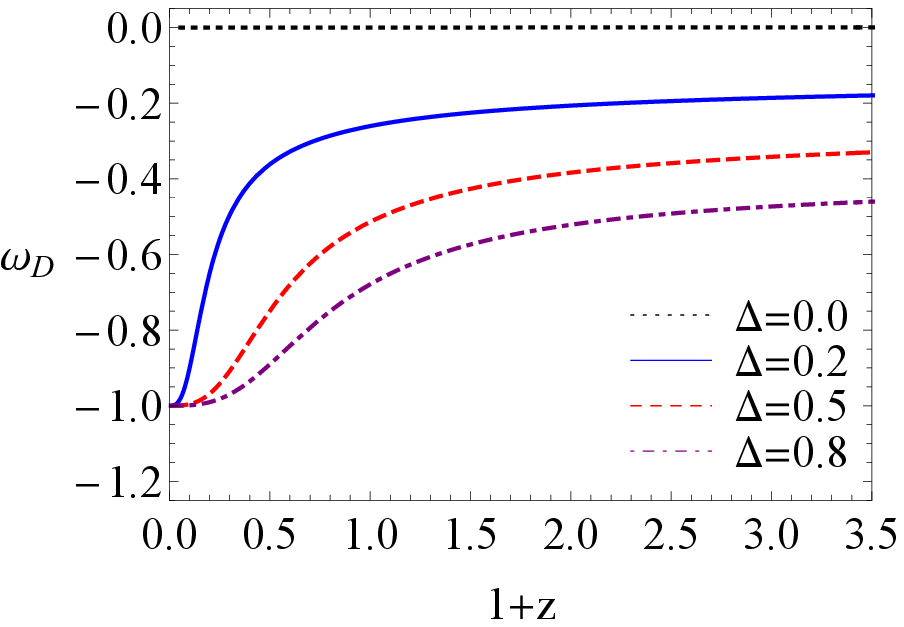}
\includegraphics[width=0.435\textwidth]{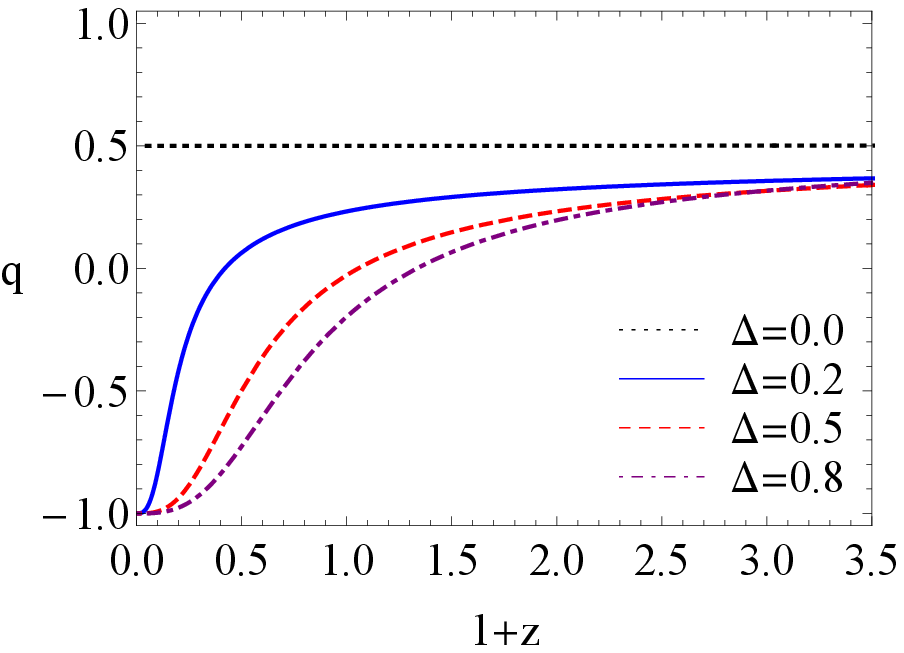}
\caption{\label{Fig11} Evolution curves of $w_{D}$ and $q$ versus redshift parameter $1+z$ for BHDEH.}
\end{center}
\end{figure*}

\begin{figure*}[htp]
\begin{center}
\includegraphics[width=0.45\textwidth]{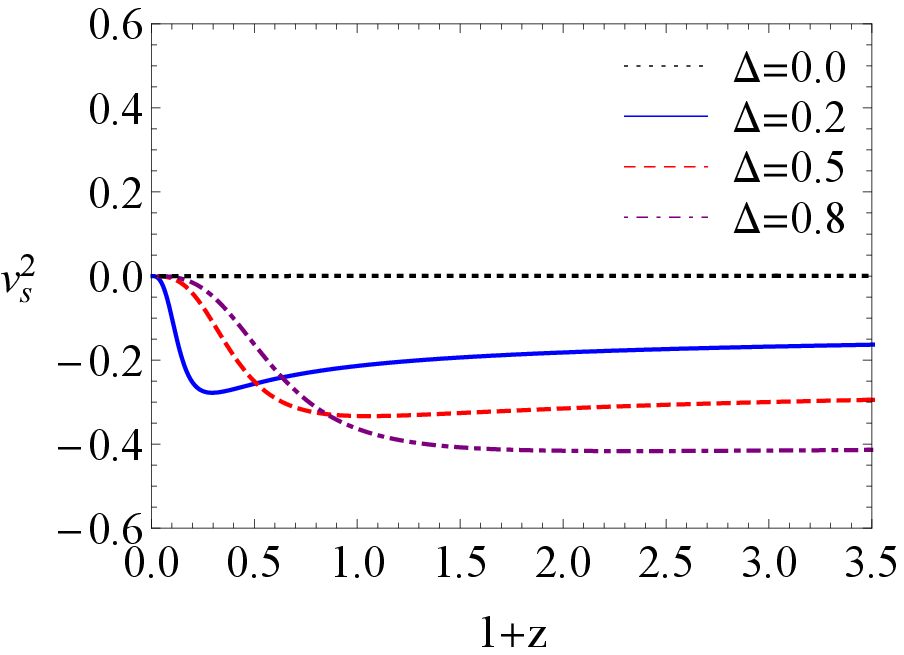}
\includegraphics[width=0.4035\textwidth]{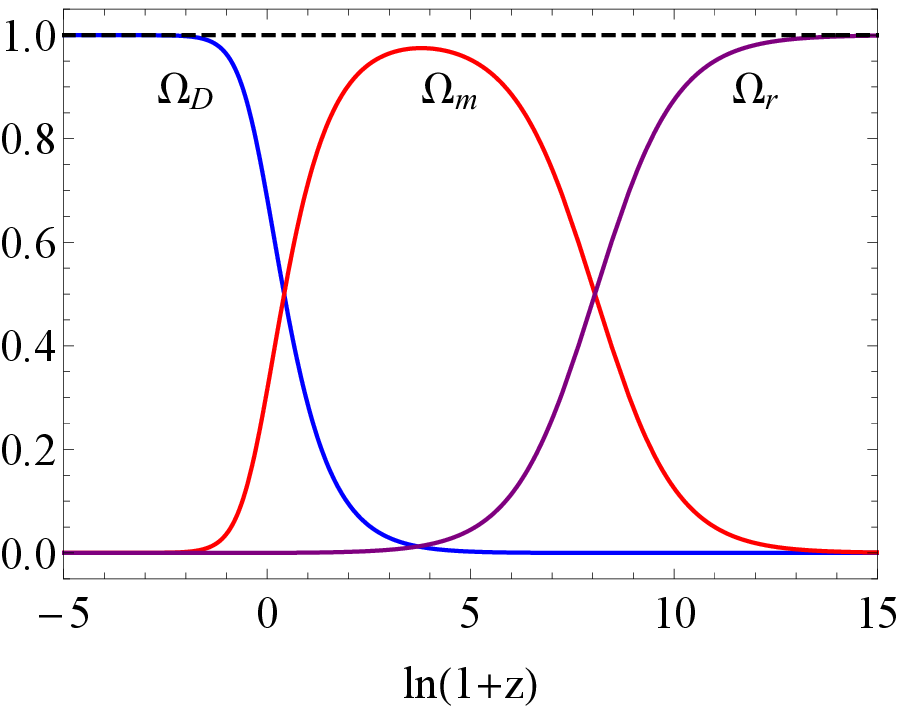}
\caption{\label{Fig12} Evolution curves of $v_{s}^{2}$ and density parameters versus redshift parameter for BHDEH. The right panel was plotted for $\Delta=0.8$.}
\end{center}
\end{figure*}

Using the initial conditions and solving Eqs.~(\ref{mD1}) and~(\ref{mD2}), we obtain the evolution curves of $\Omega_{D}$, $\Omega_{m}$, $w_{D}$, $q$ and $v_{s}^{2}$ which have been plotted in Figs.~(\ref{Fig10}), ~(\ref{Fig11}) and~(\ref{Fig12}). In Figs.~(\ref{Fig10}), ~(\ref{Fig11}) and~(\ref{Fig12}), the case $\Delta=0$ is depicted by the black dotted line. It can be seen from above figures that the current acceleration can be realized and the whole evolution of universe can described by this model. The universe can evolve into the era depicted by the standard $\Lambda$CDM model since $w_{D}$ approach to $-1$ at the late time, and a suitable range for the transition redshift is obtainable for the large value of $\Delta$. For BHDE with Hubble horizon as IR cutoff, the observable Hubble data favor a large value of $\Delta$ which can be seen in the right panel of Fig.~(\ref{Fig11ab}). Thus, for BHDE with Hubble horizon as IR cutoff, we set $\Delta=0.8$. The right panel of Fig.~(\ref{Fig12}) which is plotted for $\Delta=0.8$ shows that this model can describe the evolution of universe. However, this model is also unstable since the evolution curves of $v^{2}_{s}$ in the left panel of Fig.~(\ref{Fig12}) means $v^{2}_{s}<0$.

\subsubsection{Interacting with $Q=H(\alpha\rho_{m}+\beta\rho_{D})$}

For the situation $Q=H(\alpha\rho_{m}+\beta\rho_{D})$, we can write $w_{D}$ and $v^{2}_{s}$ as follows
\beq
w_{D}=\frac{[(\Delta-2)(\Omega_{m}+\Omega_{D}-1)-3\Delta]\Omega_{D}-2(\alpha\Omega_{m}+\beta\Omega_{D})}{[3(\Delta-2)\Omega_{D}+6]\Omega_{D}}.
\eeq
and
\bea
&&v^{2}_{s}=\frac{4(\Omega_{m}+\Omega_{D}-1)+\frac{2\alpha[(\alpha-3)\Omega_{m}+\beta \Omega_{D}]}{(\Delta-2)\Omega_{D}}}{3[(\alpha+1)\Omega_{m}+(\beta+4)\Omega_{D}-4]}\nonumber\\
&&\quad +\frac{2\Delta(\alpha+1)\Omega_{m}+4(2\Delta+\beta)(\Omega_{D}-1)}{3[(\Delta-2)\Omega_{D}+2]^{2}}
\eea

\begin{figure*}[htp]
\begin{center}
\includegraphics[width=0.45\textwidth]{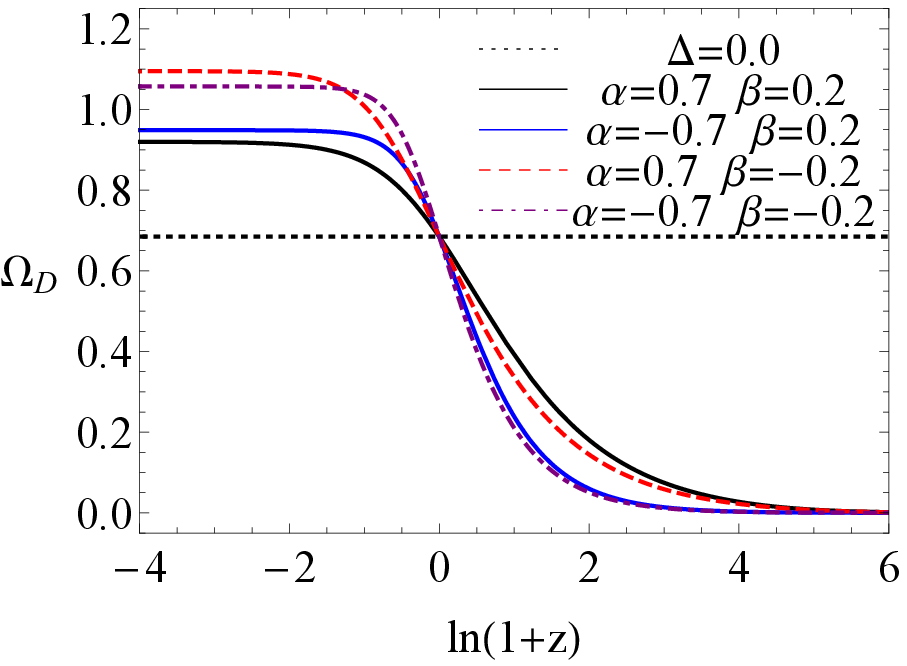}
\includegraphics[width=0.45\textwidth]{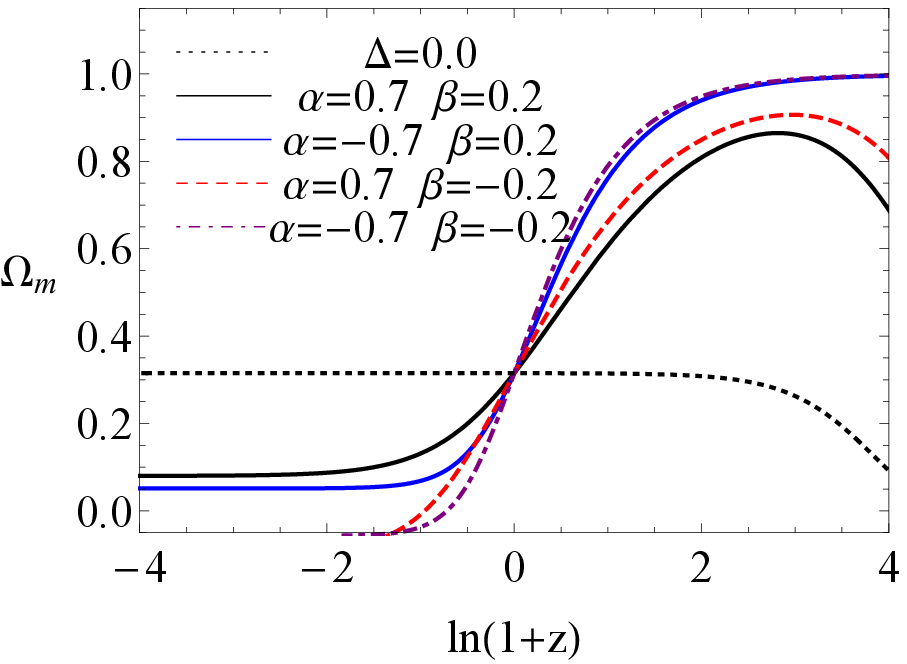}
\caption{\label{Fig13} Evolution curves of $\Omega_{D}$ and $\Omega_{m}$ versus redshift parameter $ln(1+z)$ for IBHDEHA.}
\end{center}
\end{figure*}

\begin{figure*}[htp]
\begin{center}
\includegraphics[width=0.45\textwidth]{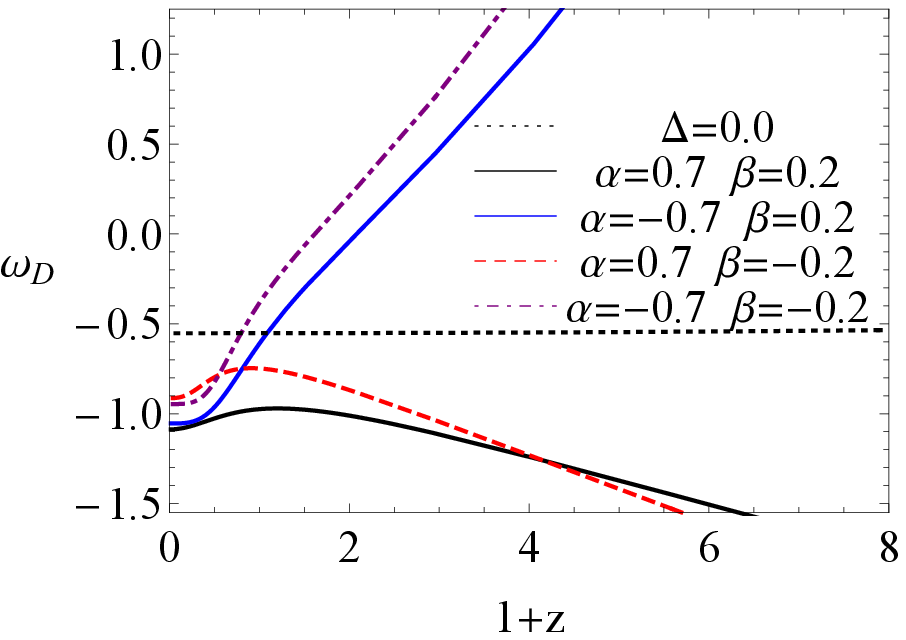}
\includegraphics[width=0.435\textwidth]{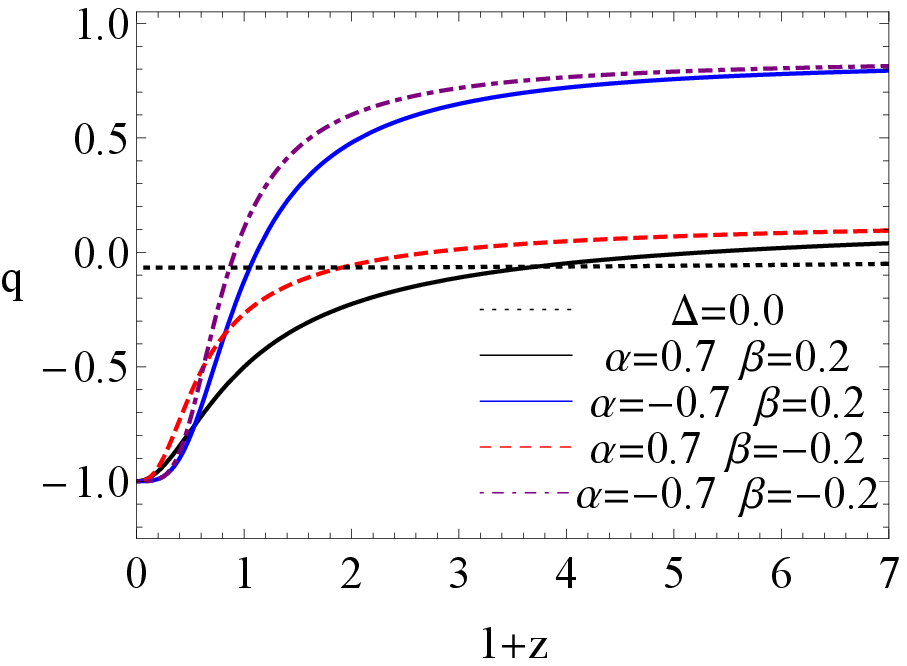}
\caption{\label{Fig14} Evolution curves of $w_{D}$ and $q$ versus redshift parameter $1+z$ for IBHDEHA.}
\end{center}
\end{figure*}

\begin{figure*}[htp]
\begin{center}
\includegraphics[width=0.45\textwidth]{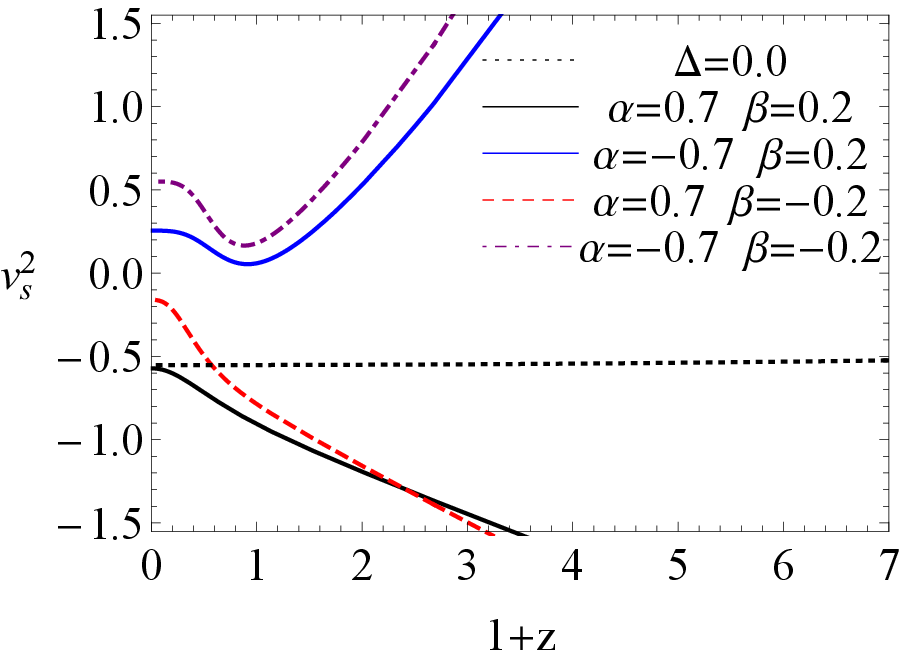}
\includegraphics[width=0.412\textwidth]{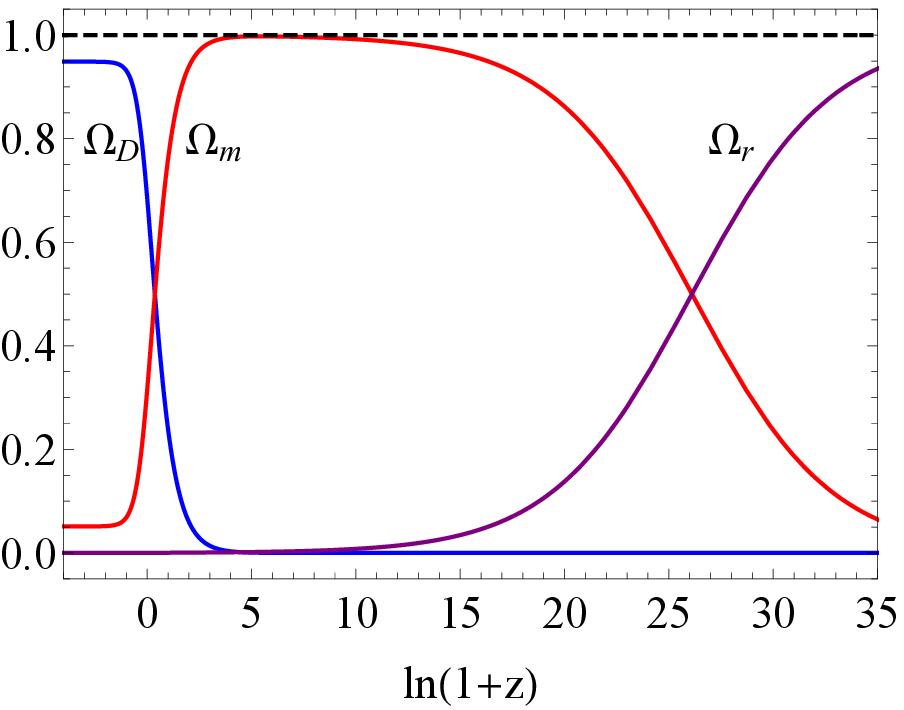}
\caption{\label{Fig15} Evolution curves of $v_{s}^{2}$ and density parameters versus redshift parameter for IBHDEHA. The right panel was plotted for $\alpha=-0.7,\beta=0.2$.}
\end{center}
\end{figure*}

In Figs.~(\ref{Fig13}), ~(\ref{Fig14}) and~(\ref{Fig15}), the evolution curves of $\Omega_{D}$, $\Omega_{m}$, $w_{D}$, $q$ and $v_{s}^{2}$ have been plotted with the initial conditions and $\Delta=0.8$. Here, we have plotted the case $\Delta=0$ with $\alpha=0.7$ and $\beta=0.2$ by the black dotted line. From Fig.~(\ref{Fig13}), we can see that $\Omega_{D}$ is larger than $1$ and the value of $\Omega_{m}$ becomes negative at the late time era for the negative $\beta$. The left panel of Fig.~(\ref{Fig15}) shows that this model can be stable for the negative $\alpha$. So, a negative $\alpha$ with a positive $\beta$ can realize the late time acceleration and ensure this model is stable. Unfortunately, the result from the right panel of Fig.~(\ref{Fig15}) shows that $\Omega_{r}$ dominates the evolution of the universe at the high redshift region $z>2\times10^{11}$ in which $a$ approaches to $0$.

\subsubsection{Interacting with $Q=\frac{\lambda}{H}\rho_{m}\rho_{D}$}

For the case $Q=\frac{\lambda}{H}\rho_{m}\rho_{D}$, after some tediously calculations, $w_{D}$ and $v^{2}_{s}$ lead to
\beq
w_{D}=\frac{[(\Delta-2)(\Omega_{m}+\Omega_{D}-1)-3\Delta]\Omega_{D}-6\lambda\Omega_{m}\Omega_{D}}{[3(\Delta-2)\Omega_{D}+6]\Omega_{D}},
\eeq
and
\bea
&&v^{2}_{s}=\frac{(\Delta-2)\Omega_{D}+(\Delta-2-6\lambda)\Omega_{m}+2-4\Delta}{3[(\Delta-2)\Omega_{D}+2]}\nonumber\\
&&\qquad +\frac{[(\Delta-2-6\lambda)\Omega_{m}-4\Delta]\Omega_{D}'}{3[(\Delta-2)\Omega_{D}+2](3\lambda \Omega_{m} \Omega_{D}+\Omega_{m}+4\Omega_{D}-4)}\nonumber\\
&&\qquad -\frac{(\Delta-2-6\lambda)\Omega_{m}'}{3(\Delta-2)(3\lambda \Omega_{m} \Omega_{D}+\Omega_{m}+4\Omega_{D}-4)}.
\eea

\begin{figure*}[htp]
\begin{center}
\includegraphics[width=0.45\textwidth]{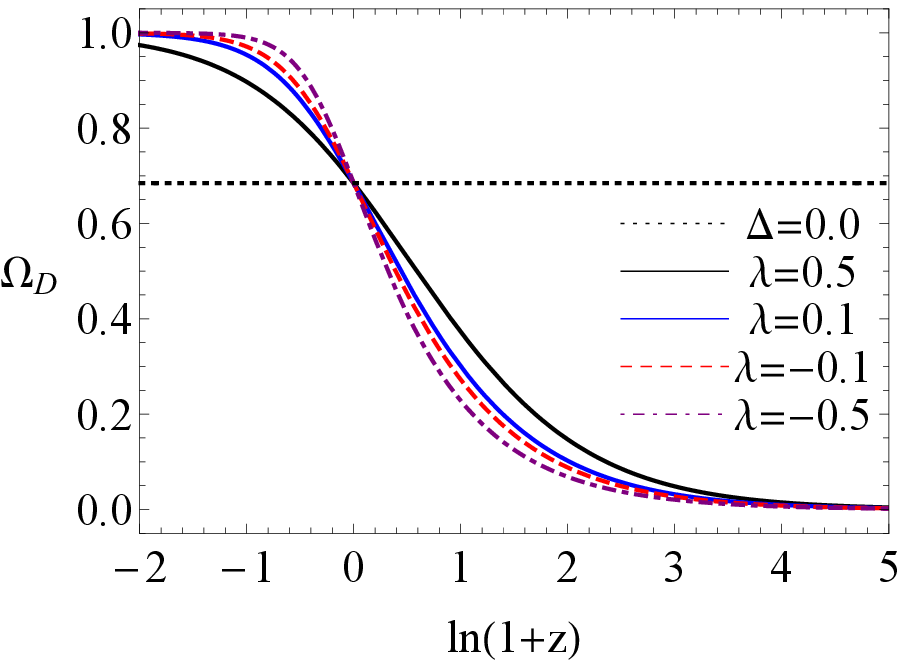}
\includegraphics[width=0.45\textwidth]{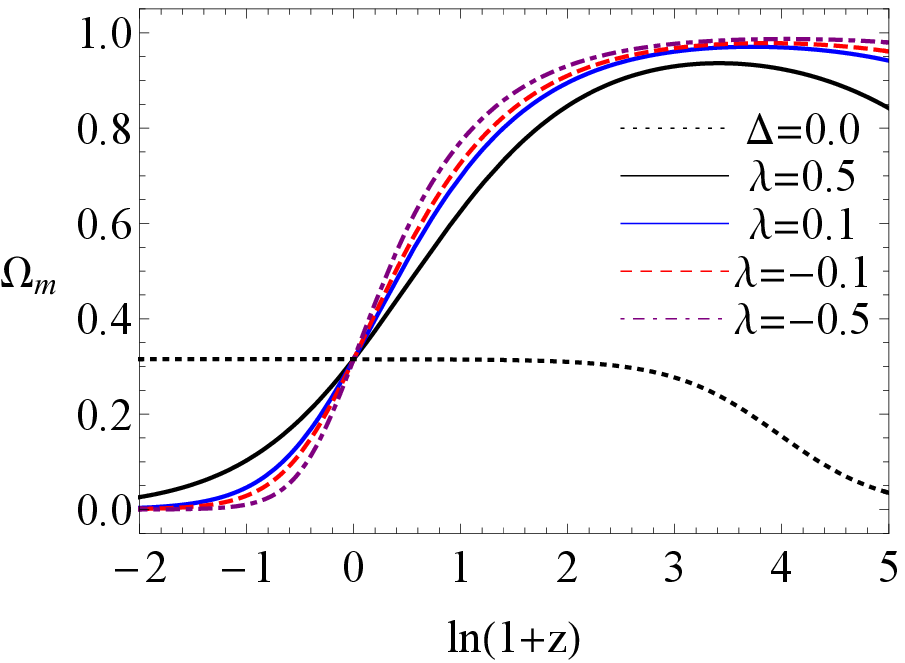}
\caption{\label{Fig16} Evolution curves of $\Omega_{D}$ and $\Omega_{m}$ versus redshift parameter $ln(1+z)$ for IBHDEHL.}
\end{center}
\end{figure*}

\begin{figure*}[htp]
\begin{center}
\includegraphics[width=0.45\textwidth]{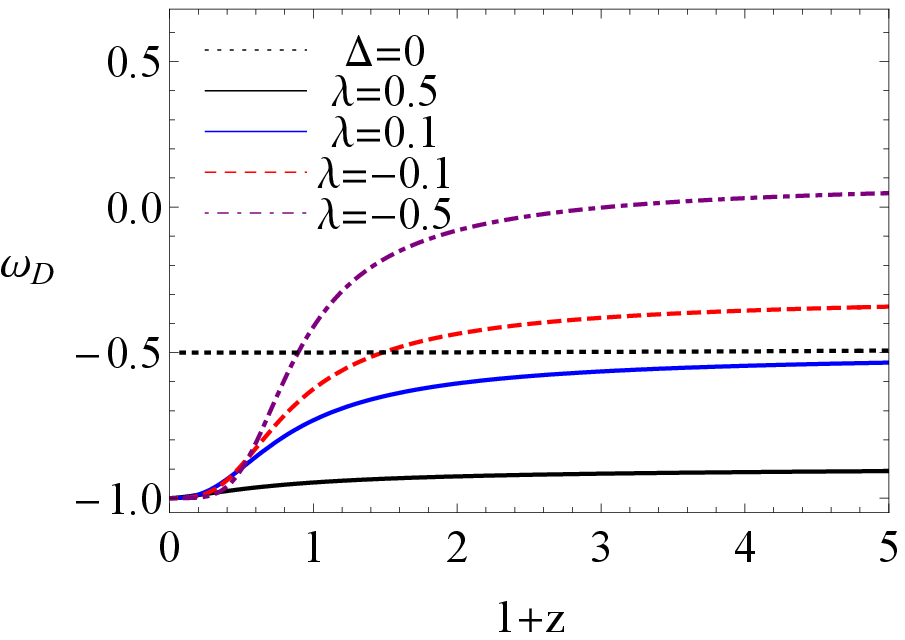}
\includegraphics[width=0.435\textwidth]{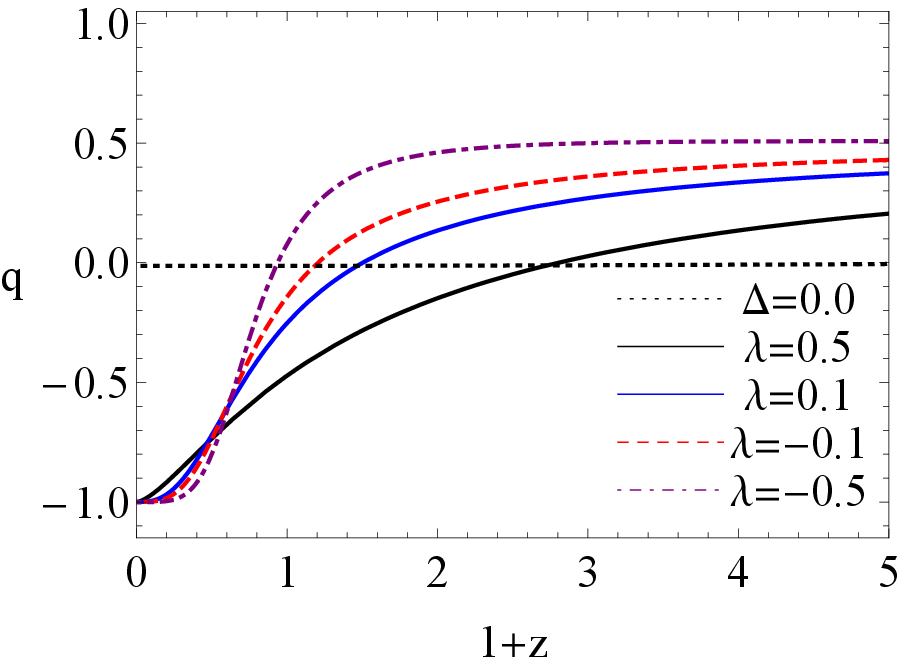}
\caption{\label{Fig17} Evolution curves of $w_{D}$ and $q$ versus redshift parameter $1+z$ for IBHDEHL.}
\end{center}
\end{figure*}

\begin{figure*}[htp]
\begin{center}
\includegraphics[width=0.45\textwidth]{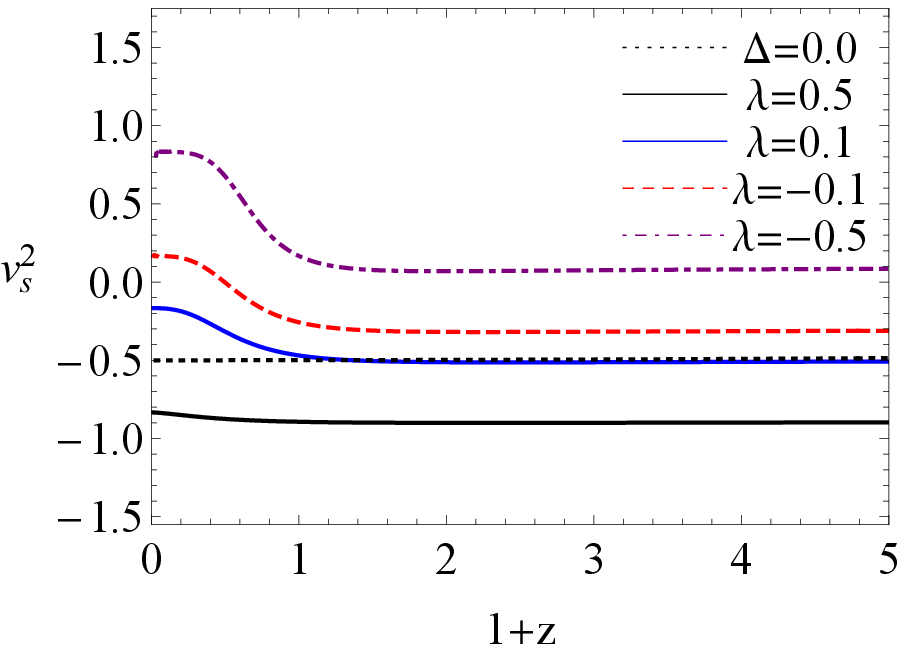}
\includegraphics[width=0.4035\textwidth]{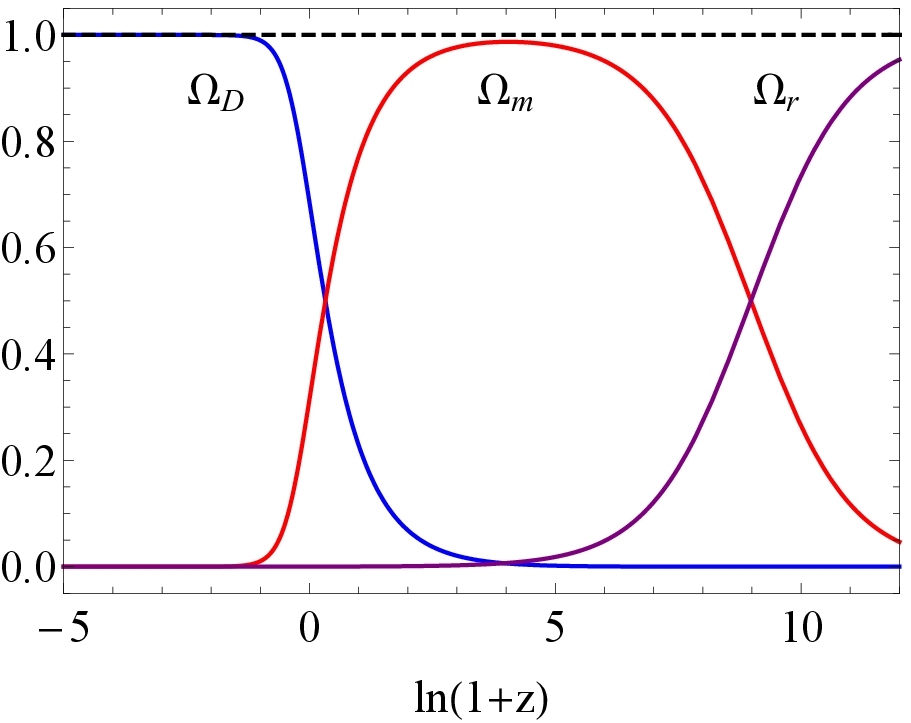}
\caption{\label{Fig18} Evolution curves of $v_{s}^{2}$ and density parameters versus redshift parameter for IBHDEHL. The right panel was plotted for $\lambda=-0.5$.}
\end{center}
\end{figure*}

The evolution curves of $\Omega_{D}$, $\Omega_{m}$, $w_{D}$, $q$ and $v_{s}^{2}$ are plotted in Figs.~(\ref{Fig16}), ~(\ref{Fig17}) and~(\ref{Fig18}) with the initial conditions and $\Delta=0.8$. The black dotted line in Figs.~(\ref{Fig16}), ~(\ref{Fig17}) and~(\ref{Fig18}) represents the case $\Delta=0$ with $\lambda=0.5$. From these figures, we can see that the current acceleration and the whole evolution of universe can be realized in this model, i.e. the universe evolves from the radiation dominated era to the pressureless matter dominated era, and then enters into the BHDE dominated era. Since both $w_{D}$ and $q$ approach to $-1$, BHDE behaves as the cosmological constant at the late time evolution, and the universe will eventually evolve into an epoch described by the standard $\Lambda$CDM model. In addition, the left panel of ~(\ref{Fig18}) shows that this model is stable for some negative coupling constant $\lambda$, while it is unstable for a positive coupling constant. These results means when the energy transfers from the pressureless matter to BHDE, this model is stable. Thus, this model can be stable under some specific conditions.

\section{Dynamical analysis of these stable models with Hubble horizon as IR cutoff}

In previous section, we have discussed the cosmic evolution and stability of BHDE models, and it is found that only IBHDEHA and IBHDEHL model can be stable against perturbations under some specific conditions. In the following, we will analyze the dynamical behaviour of these two stable models.

In order to investigate the complete asymptotic behaviour of IBHDEHA and IBHDEHL models, we will use the dynamical system techniques to study the dynamical behaviour of these models. To achieve this goal, following Ref.~\cite{Bahamonde2018, Wu2010, Dutta2017, Huang2019, Huang2021}, we can obtain the critical points by solving the autonomous system equations
\beq
\Omega_{m}'=\Omega_{D}'=0.
\eeq
By linearizing the autonomous system equations, we obtain the corresponding first order differential equations. Then, the stability of critical points can be determined by the eigenvalues of the coefficient matrix of the first order differential equations. For a critical point, if all eigenvalues are negative, it is a stable point which represents an attractor. If all eigenvalues are positive, the corresponding point is unstable. If the eigenvalues have different signs, it denotes a saddle point.

\subsection{Interacting with $Q=H(\alpha\rho_{m}+\beta\rho_{D})$}

In the case for taking the Hubble horizon as IR cutoff, the autonomous system consists of Eqs.~(\ref{mD1}) and~(\ref{mD2}). Then, solving $\Omega_{m}'=\Omega_{D}'=0$, we obtain three critical points given in Table~\ref{Tab1}. The value of $\Omega_{r}$ can be obtained by Eq.(~\ref{Orm1}). For $q<0$, the corresponding critical point represents an acceleration phase, otherwise it is a deceleration one.

Thus, point $A_{1}$ denotes the radiation dominated deceleration epoch, point $A_{2}$ represents the pressureless matter dominated deceleration epoch, and point $A_{3}$ is an acceleration epoch determined by the value of $\alpha$ and $\beta$. For the case $\beta=0$, point $A_{3}$ becomes $(0,1)$ and $w_{D}=-1$ is obtained. Thus, $A_{3}$ can behave as the cosmological constant $\Lambda$ dominated acceleration era under the condition $\beta=0$.

\begin{table*}
\caption{\label{Tab1} Critical points and the stability conditions of IBHDEHA.}
\begin{center}
 \begin{tabular}{|c|c|c|c|c|c|c|}
  \hline
  \hline
  $Label$ & \makecell[c]{$Critical \ Points$ \\$(\Omega_{m}, \Omega_{D})$} & $w_{D}$ & $q$ & $Eigenvalues$ & $Conditions$ & $Points$\\
  \hline
  $A_{1}$ & $(0,0)$ & $\frac{1-\alpha-\beta-2\Delta}{3}$ & $1$ & $(1+\alpha,2\Delta)$ & $0<\Delta<1, -1<\alpha\leq 1$ & $Unstable \ point$\\
  \hline
  $A_{2}$ & $(1,0)$ & $-$ & $\frac{1-\alpha}{2}$ & $(-1-\alpha,\frac{3-\alpha}{2}\Delta)$ & $0<\Delta<1, -1<\alpha\leq 1$ & $Saddle \ point$\\
  \hline
  $A_{3}$ & $(\frac{\beta}{3-\alpha+\beta},\frac{3-\alpha}{3-\alpha+\beta})$ & $\frac{3-\alpha+\beta}{3-\alpha}$ & $-1$ & $(-4,-\frac{(3-\alpha)(3-\alpha+\beta)\Delta}{(3-\alpha)\Delta+2\beta})$ & \makecell{$-1\leq\alpha\leq 1, -1\leq\beta\leq 0, \frac{2\alpha}{\alpha-3}<\Delta<1$ \\ or $-1\leq\alpha\leq 1, 0<\beta\leq 0, 0<\Delta<1$} & $Stable \ point$\\
  \hline
  \hline
  \end{tabular}
\end{center}
\end{table*}

To analyze the stability of these critical points, we linearize this autonomous system and then obtain the eigenvalues of the corresponding critical points. These results are also shown in Table~\ref{Tab1}. Under the constraints conditions $0<\Delta<1$ and $-1<\alpha\leq 1$, point $A_{1}$ is unstable and $P_{2}$ is a saddle point, the stability of $A_{3}$ is determined by $\alpha$ and $\beta$. The stability of these points indicates that the universe stems from the radiation dominated era $(A_{1})$ and then evolves into the pressureless matter dominated era $(A_{2})$, and eventually enters into the BHDE dominated late time acceleration epoch $(A_{3})$.

\begin{figure*}[htp]
\begin{center}
\includegraphics[width=0.45\textwidth]{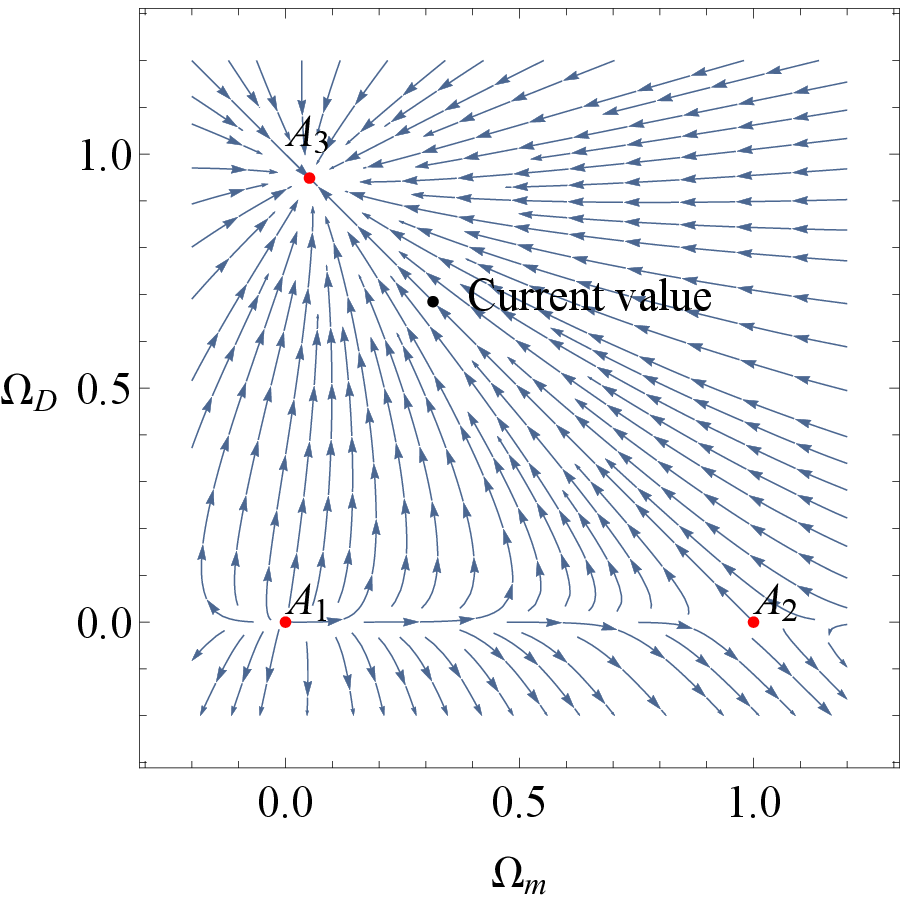}
\includegraphics[width=0.45\textwidth]{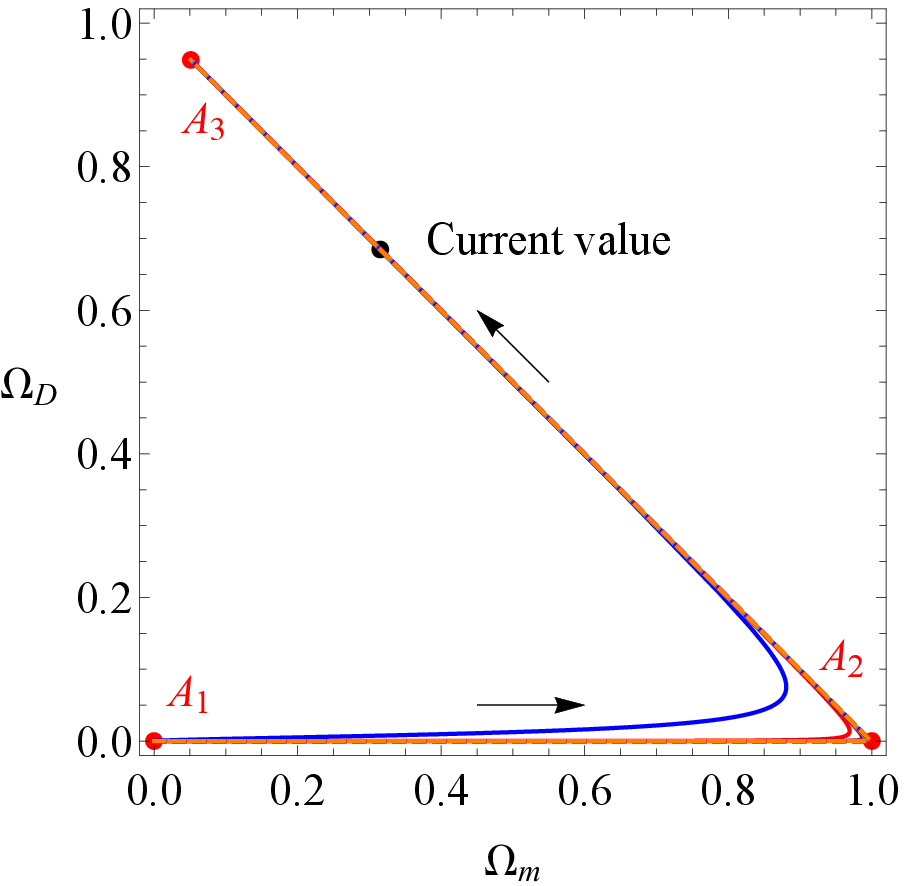}
\caption{\label{Fig7ab0} Phase space trajectories for IBHDEHA. The red points represent the critical points in Table~\ref{Tab1} and the black point denotes the current value. Here, we consider $\alpha=-0.7$ and $\beta=0.2$. The left panel is plotted for the behavior of attractor $A_{3}$ with $\Delta=0.8$, the right panel is plotted for the evolution trajectories with different value of $0<\Delta<1$.}
\end{center}
\end{figure*}

Since $A_{3}$ is a stable point with $\Omega_{m}=\frac{\beta}{3-\alpha+\beta}$ and $\Omega_{D}=\frac{3-\alpha}{3-\alpha+\beta}$, the ratio of energy density $r=\frac{\rho_{m}}{\rho_{D}}=\frac{\Omega_{M}}{\Omega_{D}}$ can approach a constant, which means the coincidence problem can be solved. An example of this case is shown in Fig.~(\ref{Fig7ab0}). The behavior of attractor $A_{3}$ is shown in the left panel of Fig.~(\ref{Fig7ab0}), and the trajectories are shown in the right one.

\begin{figure*}[htp]
\begin{center}
\includegraphics[width=0.45\textwidth]{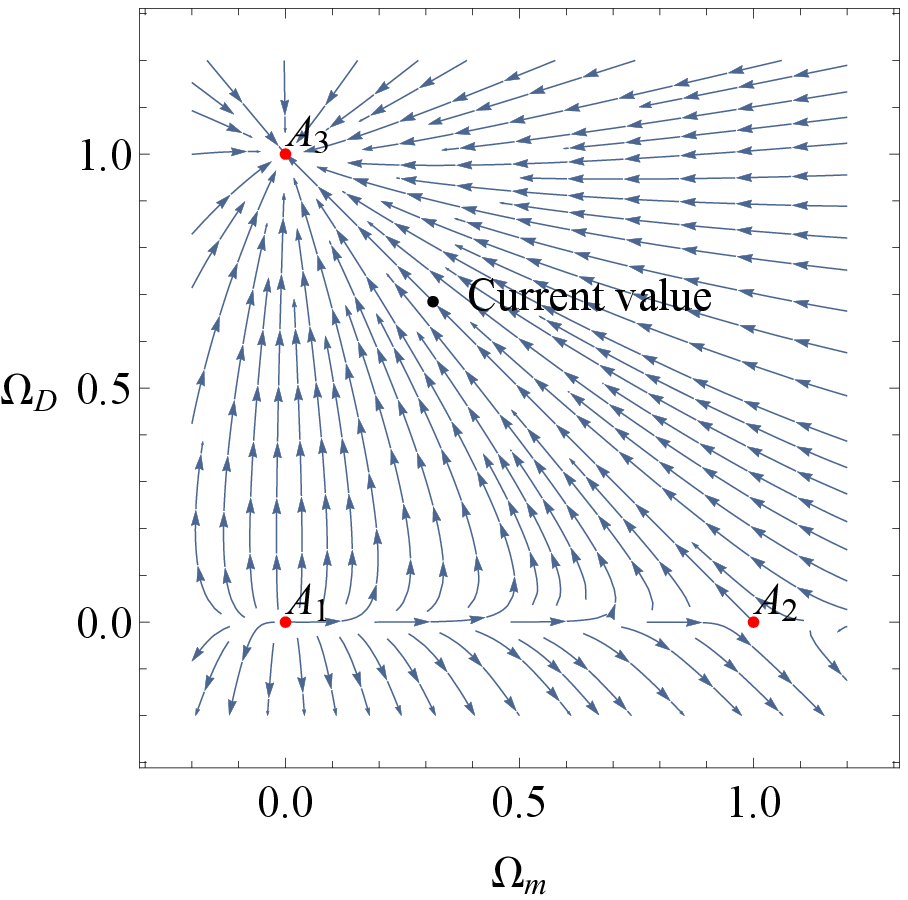}
\includegraphics[width=0.45\textwidth]{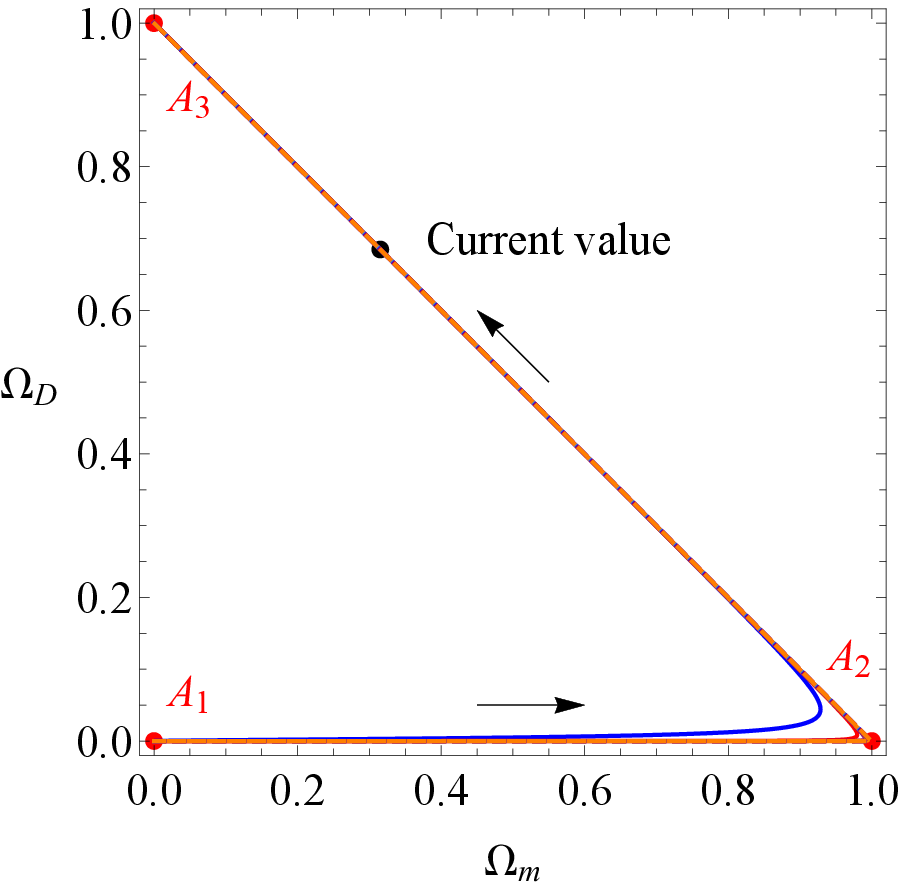}
\caption{\label{Fig7aba} Phase space trajectories for IBHDEHA with $\alpha=-0.7$ and $\beta=0$. The left panel is plotted for the behavior of attractor $A_{3}$ with $\Delta=0.8$, the right panel is plotted for the evolution trajectories with different value of $0<\Delta<1$.}
\end{center}
\end{figure*}

For the case $\beta=0$, the universe evolves from the radiation dominated era $(A_{1})$ into the pressureless matter dominated era $(A_{2})$, and eventually enters the BHDE dominated late time acceleration epoch $(A_{3})$ which can behave as the cosmological constant $\Lambda$. The phase space and evolution trajectories of this case have been plotted in Fig.~(\ref{Fig7aba}).

\begin{figure*}[htp]
\begin{center}
\includegraphics[width=0.45\textwidth]{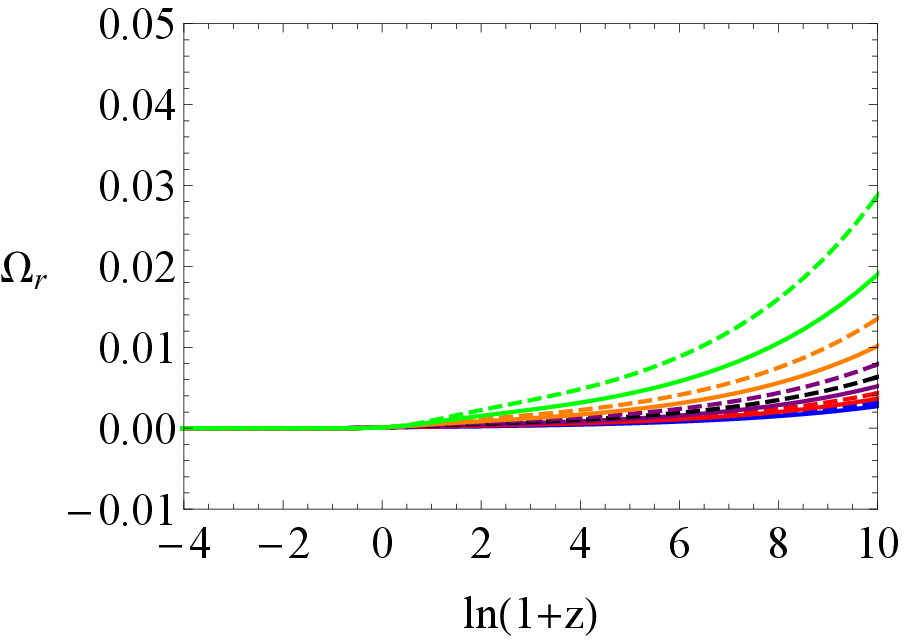}
\includegraphics[width=0.45\textwidth]{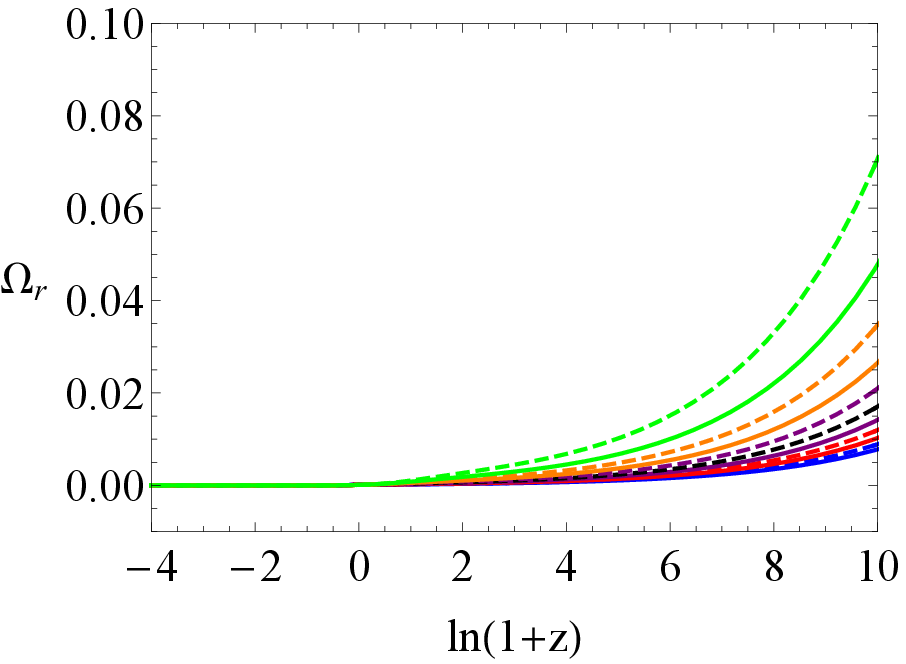}
\caption{\label{Fig7abb} Evolution curves of $\Omega_{r}$ for IBHDEHA with $-1<\beta<1$. The left panel is plotted for $\Delta=0.8$ and $\alpha=-0.7$, the right panel is for $\Delta=0.9$ and $\alpha=-0.6$.}
\end{center}
\end{figure*}

Although the IBHDEHA model is stable under some specific conditions and can describe the complete evolution epoch of the universe, $\Omega_{r}$ cannot dominate the evolution of the universe in the suitable redshift regions. Some examples are plotted in Fig.~(\ref{Fig7abb}) and the right panel of Fig.~(\ref{Fig15}). So, this model cannot describe the whole evolution history of the universe.

\subsection{Interacting with $Q=\frac{\lambda}{H}\rho_{m}\rho_{D}$}

Solving $\Omega_{m}'=\Omega_{D}'=0$, we obtain four critical points shown in Table~\ref{Tab2}. From this table, it can be seen that point $P_{1}$ denotes a deceleration epoch dominated by radiation, point $P_{2}$ represents a deceleration epoch dominated by the pressureless matter, and point $P_{3}$ denotes the BHDE dominated acceleration era. Point $P_{4}$ is fully determined by the value of $\lambda$. It is easily seen that $P_{3}$ can behave as the cosmological constant $\Lambda$ dominated acceleration era since $w_{D}=-1$ and $q=-1$.

By linearizing this autonomous system, we obtain the eigenvalues of the corresponding critical points. These results are also shown in Table~\ref{Tab2}. Under the constraint condition $0<\Delta<1$, point $P_{1}$ is unstable and $P_{2}$ is a saddle point, the stability of $P_{3}$ and $P_{4}$ are determined by $\lambda$. For $\lambda<1$, $P_{3}$ is a stable point, while a stable point $P_{4}$ requires $\lambda>1$. Although $P_{3}$ and $P_{4}$ both behave as an attractor, they cannot be stable simultaneously.

\begin{table*}
\caption{\label{Tab2} Critical points and the stability conditions of IBHDEHL.}
\begin{center}
 \begin{tabular}{|c|c|c|c|c|c|c|}
  \hline
  \hline
  $Label$ & \makecell[c]{$Critical \ Points$ \\$(\Omega_{m}, \Omega_{D})$} & $w_{D}$ & $q$ & $Eigenvalues$ & $Conditions$ & $Points$\\
  \hline
  $P_{1}$ & $(0,0)$ & $\frac{1}{3}(1-2\Delta)$ & $1$ & $(1,2\Delta)$ & $0<\Delta<1$ & $Unstable \ point$\\
  \hline
  $P_{2}$ & $(1,0)$ & $-\frac{1}{2}\Delta-\lambda$ & $\frac{1}{2}$ & $(-1,\frac{3}{2}\Delta)$ & $0<\Delta<1$ & $Saddle \ point$\\
  \hline
  $P_{3}$ & $(0,1)$ & $-1$ & $-1$ & $(-4,-3(1-\lambda))$ & $0<\Delta<1, \lambda<1$ & $Stable \ point$\\
  \hline
  $P_{4}$ & $(1-\frac{1}{\lambda},\frac{1}{\lambda})$ & $-\lambda$ & $-1$ & $(-4,-\frac{3\Delta(1-\lambda)}{2(1-\lambda)-\Delta})$ & $0<\Delta<1, \lambda>1$ & $Stable \ point$\\
  \hline
  \hline
  \end{tabular}
\end{center}
\end{table*}

\begin{figure*}[htp]
\begin{center}
\includegraphics[width=0.45\textwidth]{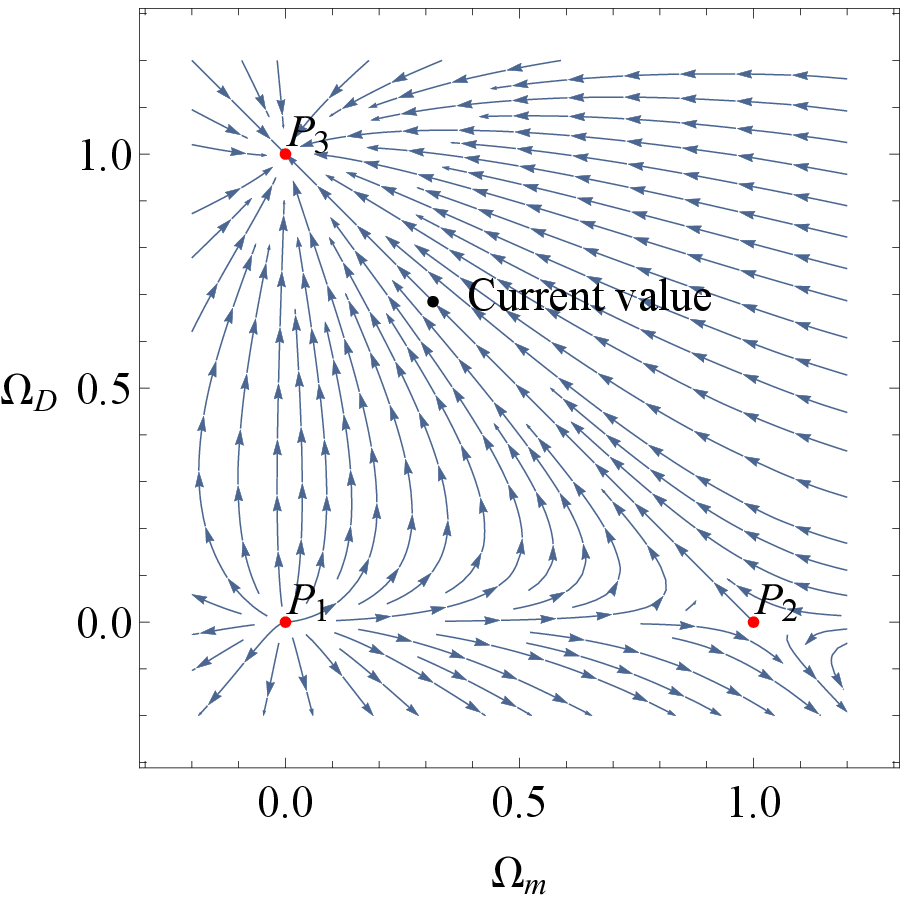}
\includegraphics[width=0.45\textwidth]{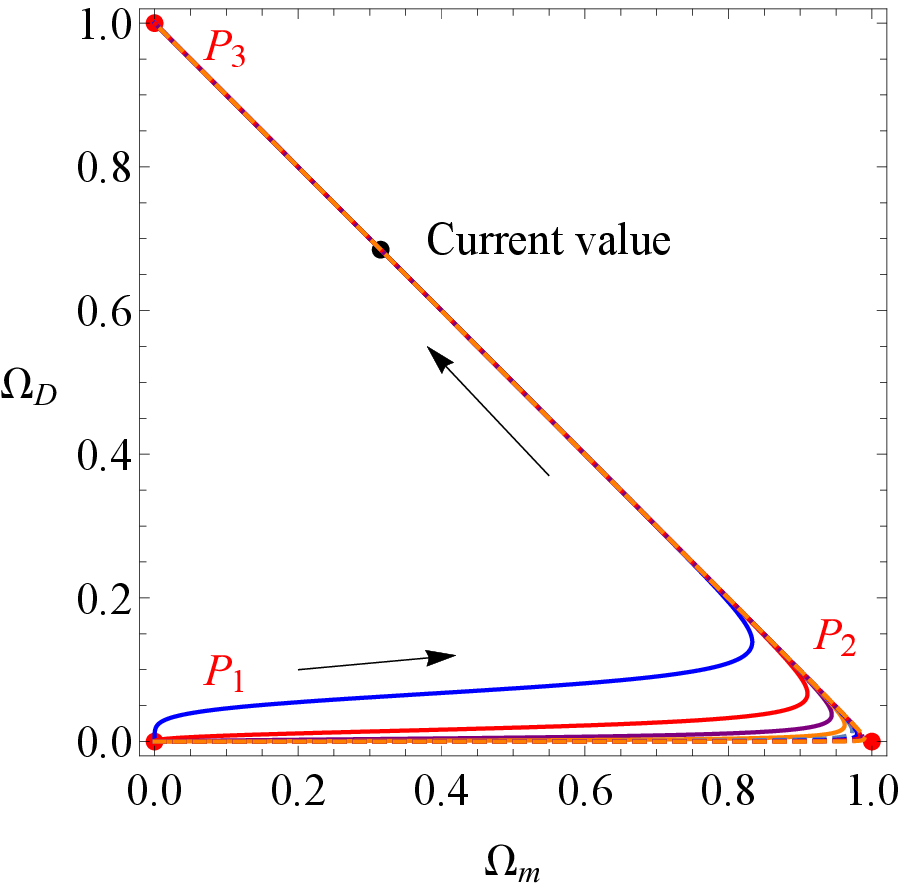}
\caption{\label{Fig7ab} Phase space trajectories for IBHDEHL with $\lambda=-0.5$. The left panel is plotted for the behavior of attractor $P_{3}$ with $\Delta=0.8$, the right panel is plotted for the evolution trajectories with different value of $0<\Delta<1$.}
\end{center}
\end{figure*}

Thus, for $\lambda<1$, these results show that the universe evolves from the radiation dominated era $(P_{1})$ into the pressureless matter dominated era $(P_{2})$, and eventually enters the BHDE dominated late time acceleration epoch $(P_{3})$ which can behave as the cosmological constant $\Lambda$. Some examples of this situations are plotted in Fig.~(\ref{Fig7ab}). The left panel describes the behavior of attractor $P_{3}$. And the right panel depicts the phase space trajectories on the $\Omega_{m}-\Omega_{D}$ plane in which all trajectories stem from $P_{1}$, approach to $P_{2}$, and eventually converge into $P_{3}$. In this situation, the whole history of the universe can be described by this model.

\begin{figure*}[htp]
\begin{center}
\includegraphics[width=0.45\textwidth]{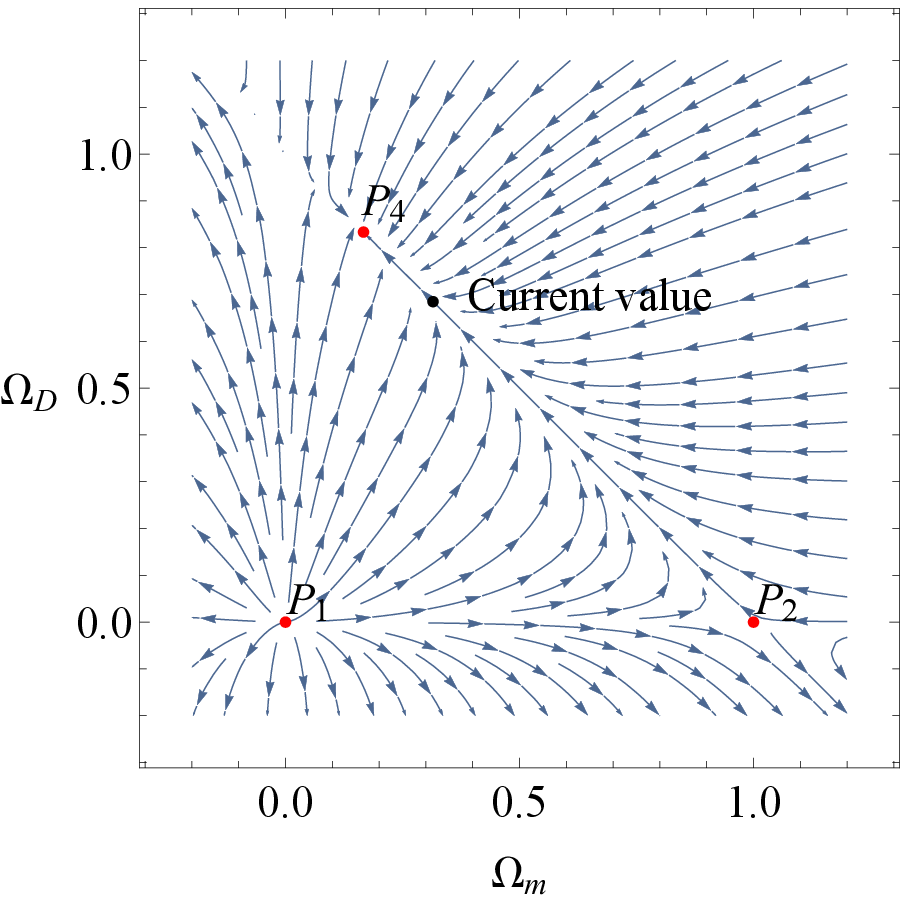}
\includegraphics[width=0.45\textwidth]{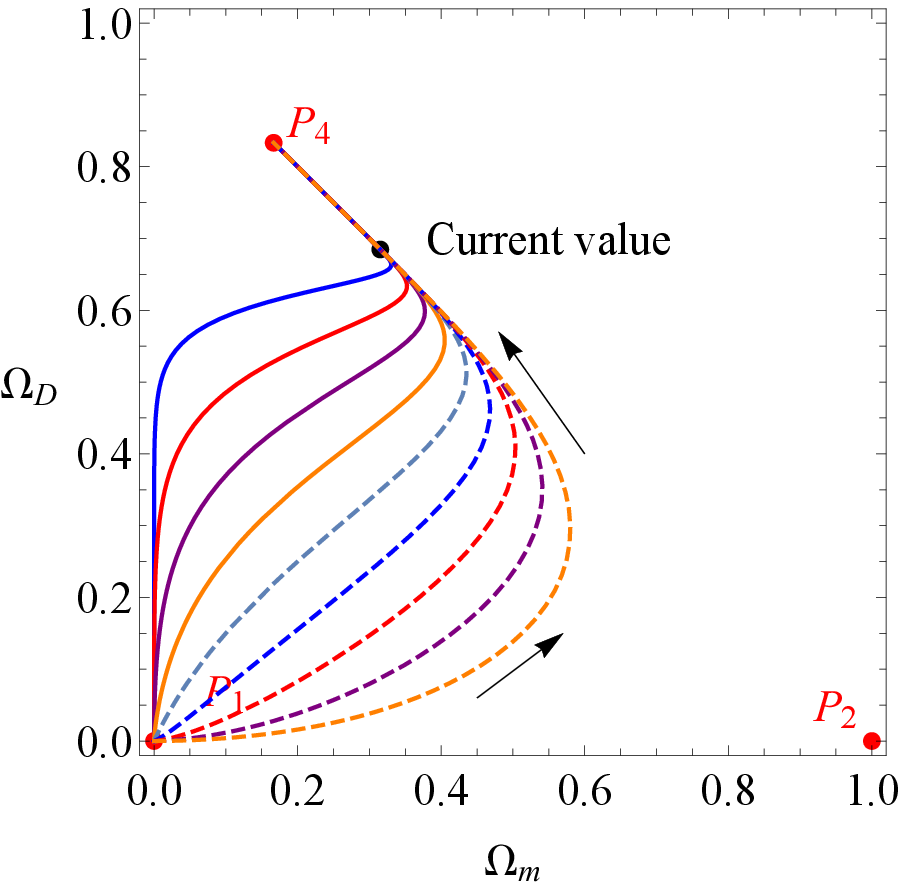}
\caption{\label{Fig7cd} Phase space trajectories for IBHDEHL with $\lambda=1.2$. The left panel is plotted for the behavior of attractor $P_{4}$ with $\Delta=0.8$, the right panel is plotted for the evolution trajectories with different value of $0<\Delta<1$.}
\end{center}
\end{figure*}

For $1<\lambda<2$, the stability of these points indicates that the universe stems from the radiation dominated era $(P_{1})$ and then evolves close to the pressureless matter dominated era $(P_{2})$, and eventually enters into the BHDE dominated late time acceleration epoch $(P_{4})$ in which the coincidence problem can be solved. An example of this case is shown in Fig.~(\ref{Fig7cd}). The behavior of attractor $P_{4}$ is shown in the left panel of Fig.~(\ref{Fig7cd}), and the trajectories is shown in the right one. In this case, the coincidence problem is solved.

For $\lambda\geq2$, the universe will finally evolve into an accelerated epoch dominated by the pressureless matter rather than the BHDE. Although the coincidence problem can be solved in this case, the acceleration in this case is caused by the pressureless matter.

\section{Statefinder analysis}

In previous section, we have discussed the dynamical evolution of the universe in IBHDEHA and IBHDEHL models. We find only IBHDEHL model can explain the current accelerated expansion and describe the whole evolution of the universe. And an attractor behaving as the cosmological constant $\Lambda$ exists in this model. Then, how to discriminate this model from the standard $\Lambda$CDM model becomes another problem. To solve this problem, we use the statefinder diagnostic pairs $\{r, s\}$, which was introduced by Sahni \textit{et al.}~\cite{Sahni2003}, to analyze this model.

The statefinder parameters $r$ and $s$ are defined as~\cite{Sahni2003}
\beq\label{rs}
r=\frac{\dddot{a}}{a H^{3}}, \qquad s=\frac{r-1}{3(q-\frac{1}{2})}.
\eeq
The statefinder parameter is a geometrical diagnostic since it only depends on $a$. By differentiating Eq.~(\ref{HH2}), we can express the statefinder parameters $r$ and $s$ by $\Omega_{m}'$ and $\Omega_{D}'$. Then, solving Eqs.~(\ref{mD1}) and~(\ref{mD2}) numerically, we can obtain the evolution curves of the universe in the parameter space $r-s$.

\begin{figure*}[htp]
\begin{center}
\includegraphics[width=0.45\textwidth]{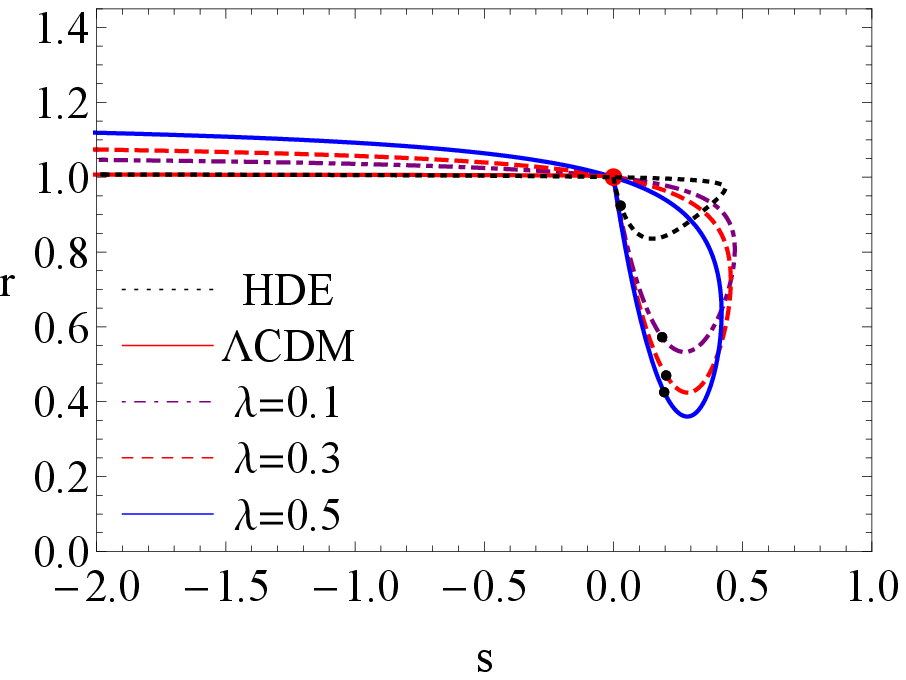}
\includegraphics[width=0.45\textwidth]{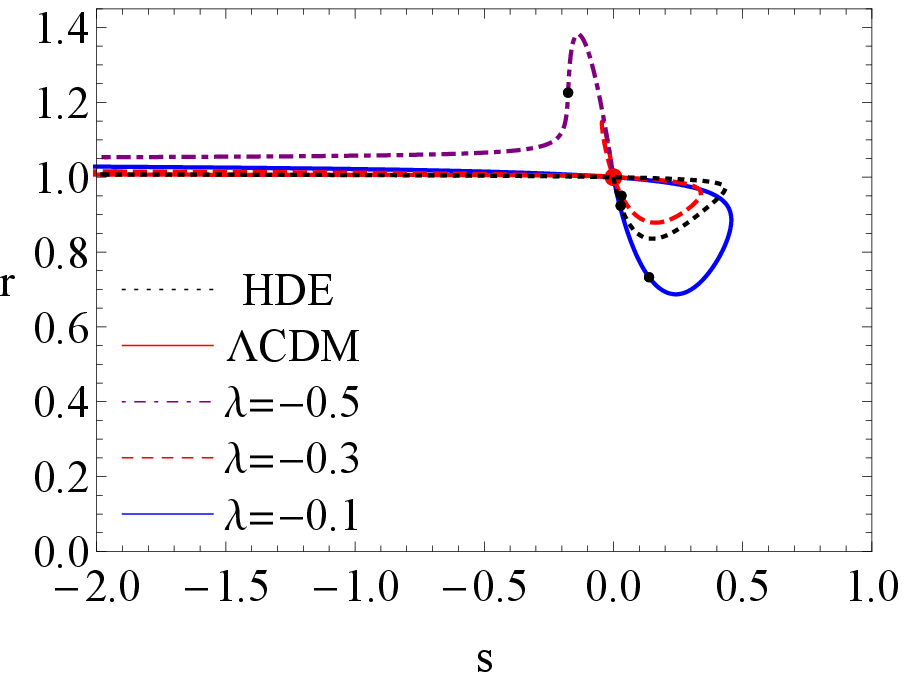}
\caption{\label{Fig8ab} Illustrative examples for the statefinder diagnostic $\{r, s\}$ of IBHDEHL. Present values of $\{r, s\}$ are marked by black dots, while the present values of the $\Lambda$CDM model is a fixed point $\{r, s\}=\{1, 0\}$ marked by the red dot. Here, we take $\Delta=0.8$.}
\end{center}
\end{figure*}

\begin{figure*}[htp]
\begin{center}
\includegraphics[width=0.45\textwidth]{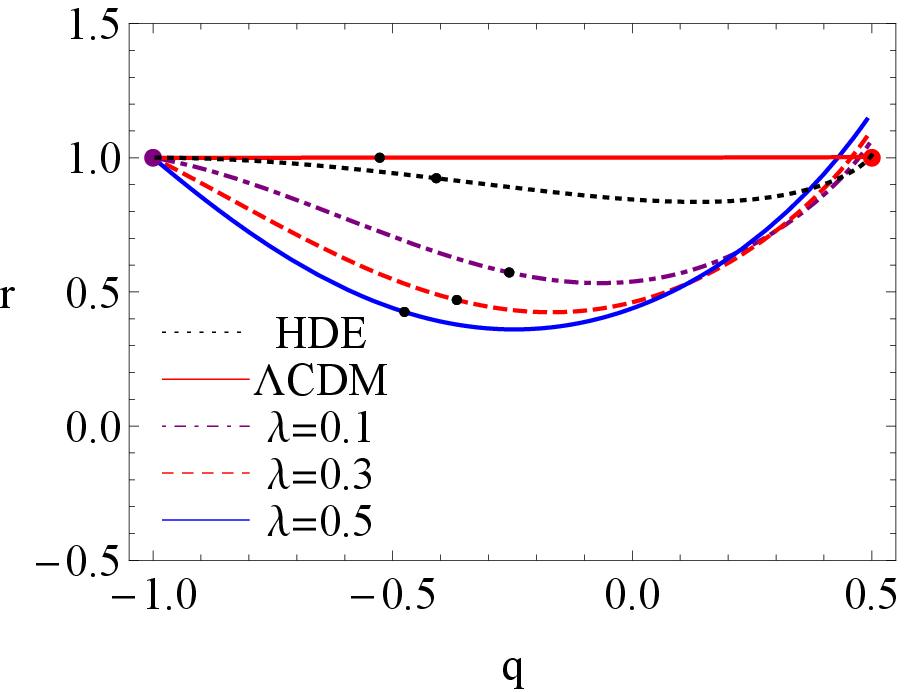}
\includegraphics[width=0.44\textwidth]{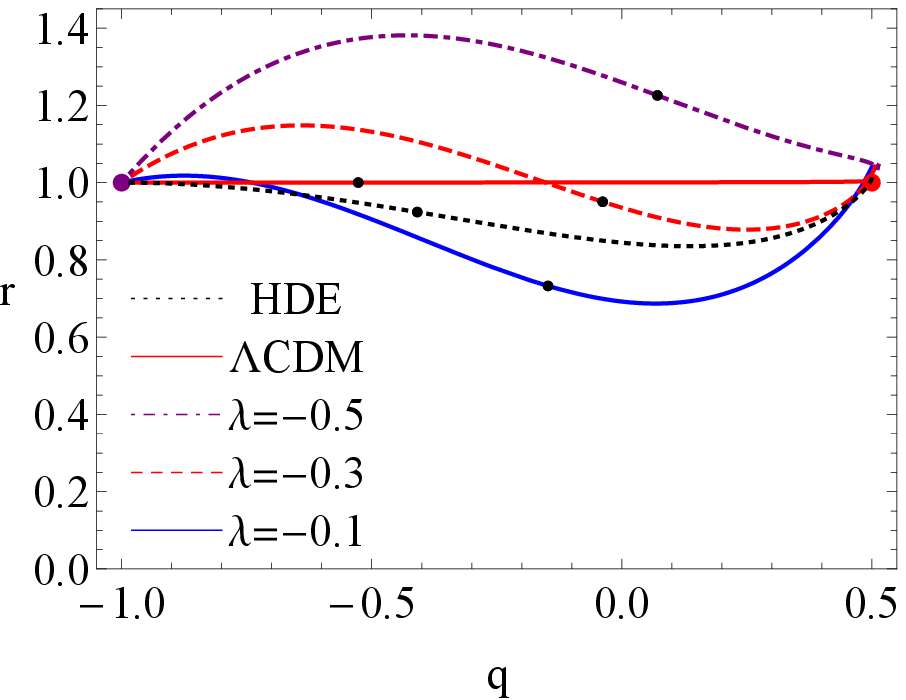}
\caption{\label{Fig9ab} Illustrative examples for the statefinder diagnostic $\{r, q\}$ of IBHDEHL. Present values of $\{r, q\}$ are marked by black dots, the de Sitter expansion fixed point marked by the purple dot and the standard cold dark matter fixed point marked by the red dot. Here, we take $\Delta=0.8$.}
\end{center}
\end{figure*}

In Fig.~(\ref{Fig8ab}), we have plotted an example for the evolution curves of the statefinder diagnostic pairs $\{r, s\}$. In the left panel, we take some positive $\lambda$, and use the negative one in the right panel. The black dotted line in Fig.~(\ref{Fig8ab}) represents HDE model. From Fig.~(\ref{Fig8ab}), one can see that the statefinder pairs $\{r, s\}$ start from the left side of the $\Lambda$CDM fixed point$(0,1)$, i.e. $s<0$ and $r>1$ which is the characteristic of Chaplygin gas, and then enter into the region $s>0, r<1$ where the trajectories of quintessence and phantom stay in these regions. After passing through these regions, they finally tend to the $\Lambda$CDM fixed point$(0,1)$. It is noticeable that the behaviour of the statefinder pairs $\{r, s\}$ in these model is similar to the quintom dark energy~\cite{Wu2005} and interacting vacuum energy model~\cite{Panotopoulos2020}. In addition, for $\lambda=-0.5$, the statefinder pairs start from the Chaplygin gas region and then approach to the $\Lambda$CDM fixed point. For $\lambda=-0.3$, the present value of $\{r, s\}$ is close to the $\Lambda$CDM fixed point, and it is not for the other cases. These trajectories in $r-s$ plane demonstrate the contrasting behavior of this model and the standard $\Lambda$CDM model strikingly.

For the sake of complementarity, we have plotted the evolution trajectories for another statefinder diagram $\{r, q\}$ in Fig.~(\ref{Fig9ab}). From Fig.~(\ref{Fig9ab}), one can see that the standard $\Lambda$ model starts from the standard cold dark matter fixed point $(0.5,1)$, while the evolution curves for IBHDEHL have a smaller deviation from this fixed point. And both models can evolve into the de Sitter expansion fixed point $(-1,1)$ in the future.

In addition, under the condition $\Omega_{r}=0$ and $\Omega_{D}=0.73$, the evolution curves of HDE in Fig.~(\ref{Fig8ab}) and Fig.~(\ref{Fig9ab}) become that in~\cite{Zhang2005a}.

\section{Turning point in Hubble diagram}

\begin{figure*}[htp]
\begin{center}
\includegraphics[width=0.45\textwidth]{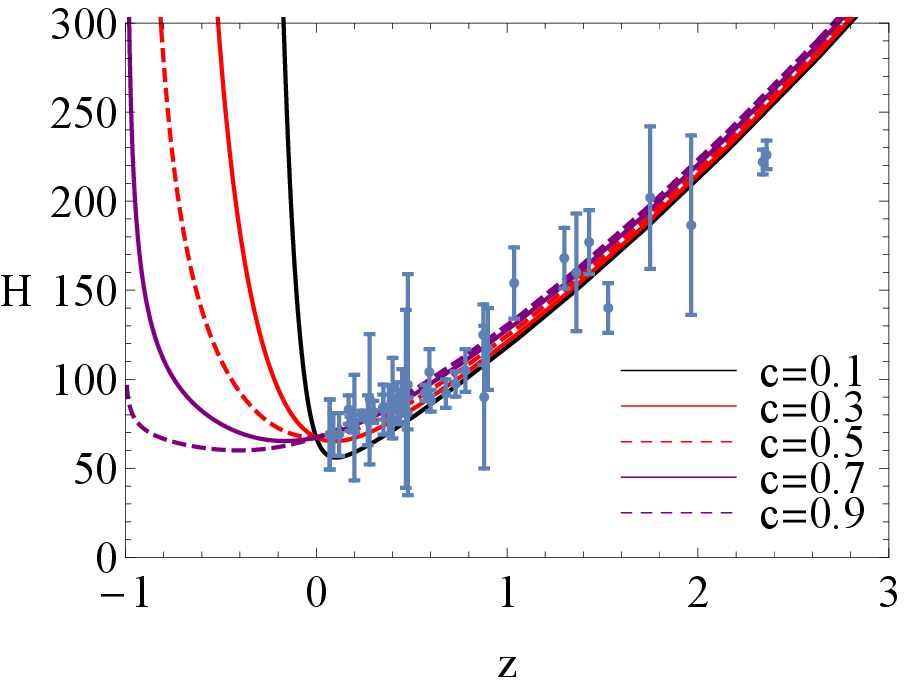}
\includegraphics[width=0.44\textwidth]{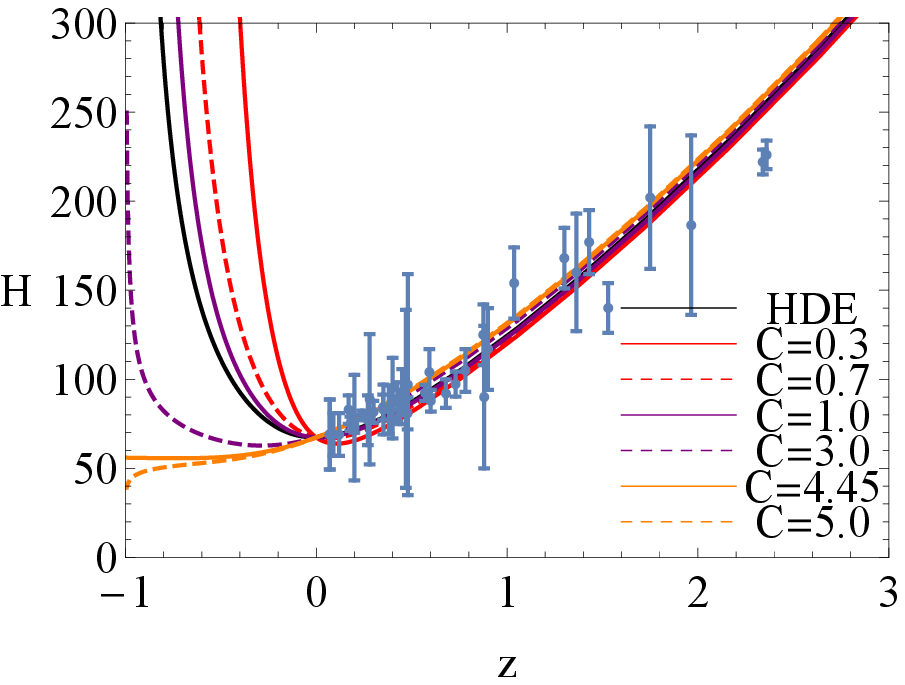}
\caption{\label{Fig10ab} Evolution curves of $H$. The left panel was plotted for HDE, while the right one was shown for BHDEF with $\Delta=0.1$.}
\end{center}
\end{figure*}

In HDE, it was found that there exists a turning point in the Hubble diagram $H(z)$, which leads to HDE model conflicting with the cosmological paradigm~\cite{Colgain2021}. This turning point is determined by the constant parameter $c$ in HDE model, and this point occurs in the future for $0.5\leq c <1$, while it becomes observable for $c\leq 0.5$. This result was obtained in~\cite{Colgain2021} and it was also shown in the left panel of Fig.~(\ref{Fig10ab}). When the cosmological observational data are used to constrain the free parameter $c$ in HDE model, the combination of different data provides different best fitting values of $c$, and some examples are shown in Table~\ref{Tab3} where most results indicate that the turning point can be postponed to the future.

\begin{table*}
\caption{\label{Tab3} Examples for best fitting values of $c$ in HDE model.}
\begin{center}
 \begin{tabular}{|c|c|c|c|c|c|c|}
  \hline
  \hline
  $Data$ & $c$ & $References$\\
  \hline
  $H(z)+SNIa+CMB+BAO$ & $0.88^{+0.21}_{-0.15}$ & ~\cite{Feng2007}\\
  \hline
  $Planck+WP+BAO+HST+lensing$ & $0.495\pm 0.039$ & ~\cite{Li2013}\\
  \hline
  $H(z)+SN+CMB+BAO$ & $0.7331^{+0.0354}_{-0.0421}$ & ~\cite{Feng2016}\\
  \hline
  $H(z)+SNIa+CMB+BAO+BBN$ & $0.785^{+0.042}_{-0.056}$ & ~\cite{Akhlaghi2018}\\
  \hline
  $Planck+BAO+R19$ & $0.51\pm 0.02$ & ~\cite{Da2020}\\
  \hline
  $CMB+BAO+SNE$ & $0.621\pm 0.026$ & ~\cite{ Colgain2021}\\
  \hline
  \hline
  \end{tabular}
\end{center}
\end{table*}

\begin{figure*}[htp]
\begin{center}
\includegraphics[width=0.45\textwidth]{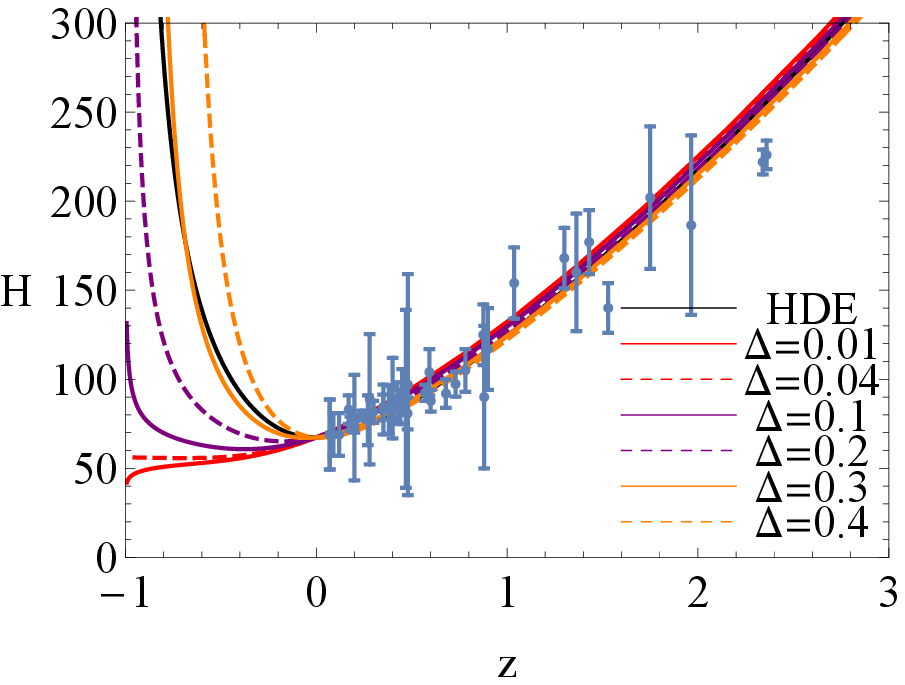}
\includegraphics[width=0.44\textwidth]{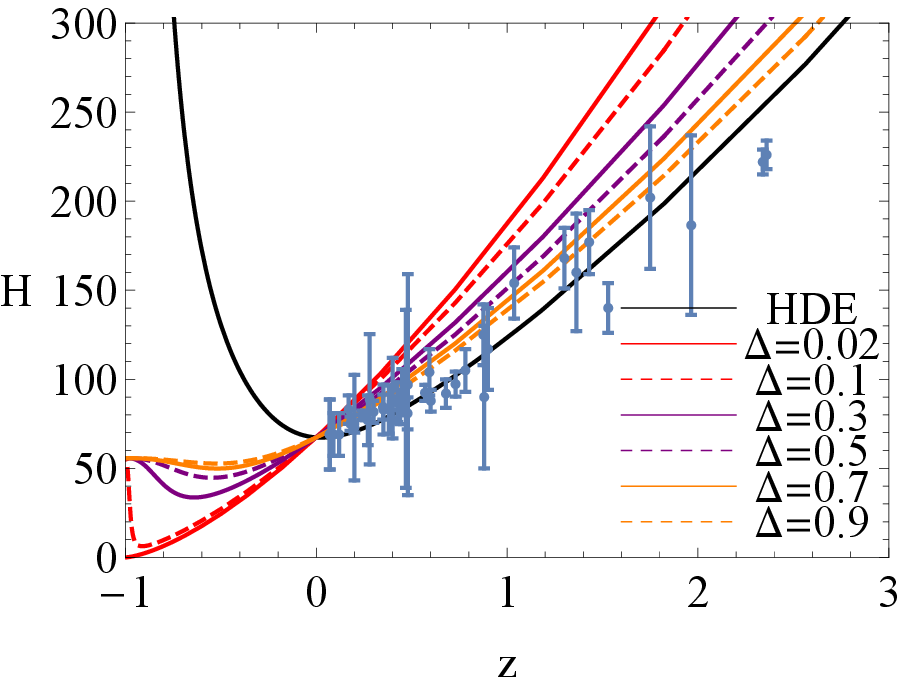}
\caption{\label{Fig11ab} Evolution curves of $H$. The left panel was plotted for BHDEF with $C=3.5$, while the right one was shown for BHDEH.}
\end{center}
\end{figure*}

Then, in order to discuss the turning point in BHDE models, we have plotted the evolution curves of Hubble parameter $H$ in Fig.~(\ref{Fig10ab}) and ~(\ref{Fig11ab}). The error bar in Fig.~(\ref{Fig10ab}) and ~(\ref{Fig11ab}) represent the observational Hubble parameter data~\cite{Akhlaghi2018, Cao2021}. In the right panel of Fig.~(\ref{Fig10ab}), the evolution curve of $H$ for BHDEF model was plotted with $\Delta=0.1$. The results show that the turning point moves to lower redshift with the increasing of $C$, and vanishes when $C \geq 4.45$. The evolution curve of $H$ for BHDEF model with $C=3.5$ was plotted in the left panel of Fig.~(\ref{Fig11ab}) which shows that this turning point moves to lower redshift regions with the decreasing of $\Delta$. For $\Delta \leq 0.04$, the turning point disappears. Recently, the combination $H(z)+SNIa$ place constraints on $C=3.421^{+1.753}_{-1.611}$ and $\Delta=0.094^{+0.094}_{-0.101}$~\cite{Anagnostopoulos2020}. Thus, under the constraints $C=3.421^{+1.753}_{-1.611}$ and $\Delta=0.094^{+0.094}_{-0.101}$, the turning point of Hubble diagram in BHDEF is nonexistent.

The evolution curves of $H$ for BHDEH with different $\Delta$ are shown in the right panel of Fig.~(\ref{Fig11ab}). This figure shows that the turning point occurs in the future, and it vanishes with the increasing of $\Delta$. For $\Delta>0.7$, the turning point can be avoided. And for $\Delta\sim 0.1$, the turning point occurs near $z=-1$ which means $a\rightarrow \infty$. When $\Delta\leq 0.02$, this point disappears. Thus, in the region $\Delta > 0.7$ or $\Delta \leq 0.02$, the turning point of Hubble diagram does not exist in BHDEH.

Recently, considering BHDE with the apparent horizon as IR cutoff (BHDEA), the Big Bang Nucleosynthesis (BBN) place a constraint on the parameter of Barrow entropy $\Delta\leq 1.4\times10^{-4}$ in order not to spoil the BBN epoch~\cite{Barrow2021}. Under this constraint, BHDEA model can return to HDE model and the turning point in Hubble diagram cannot be avoided. In BHDEF model, the combination $H(z)+SNIa$ give looser constraints $\Delta=0.094^{+0.094}_{-0.101}$ and $C=3.421^{+1.753}_{-1.611}$, which provide us an approach to solve the turning point in Hubble diagram. In BHDEA and BHDEF models, the constraints indicate that a small $\Delta$ is favored by the current observations. However, in BHDEH model, the right panel of Fig.~(\ref{Fig11ab}) indicates the observable Hubble data favor $\Delta>0.7$, which is different from the values of $\Delta$ in BHDEA and BHDEF models. To obtain a best fitting value of the parameters in BHDEA, BHDEF and BHDEH models, one require to combine more observational data to constrain the parameters. It is interesting to compare BHDEA, BHDEF and BHDEH models and combine more current observations to constrain the parameters in these models, and this is what we shall do in the future.

\section{Conclusion}

Based on the holographic principle and the Barrow entropy, which results from the modification of the black hole surface due to quantum-gravitational effects, a new HDE model named BHDE has been proposed. In this paper, by considering the future event horizon and the Hubble horizon as IR cutoff, we analyze the evolution and stability of BHDE models with different interaction terms. We find all of these BHDE models can explain the current accelerated expansion, but only IBHDEHA and IBHDEHL can be stable under some specific conditions in which the coupling constant requires to be negative. A negative coupling constant may result in $Q<0$, which indicates that the pressureless matter is giving energy to BHDE. When we study the evolution of the universe in IBHDEHA and IBHDEHL models, we find only IBHDEHL model can describe the whole evolution of the universe, i.e. the universe stems from the radiation dominated epoch, passes through the pressureless matter dominated epoch, and eventually approaches the dark matter dominated acceleration epoch. Since both $w_{D}$ and $q$ can approach to $-1$ at the late time evolution of the universe, the BHDE in IBHDEHL model can behave as the cosmological constant.

Then, we apply the dynamical analysis techniques to IBHDEHA and IBHDEHL models which are stable against perturbations. Although IBHDEHA model can describe the entire evolution epoch of the universe, $\Omega_{r}$ cannot dominate the evolution of the universe in the corresponding redshift regions. Thus, only IBHDEHL model can describe the whole evolution of the universe. In this model, the results show that there exist three different critical points for $\lambda<1$, namely the unstable point represents the radiation dominated epoch, the saddle point denotes the pressureless matter dominated epoch, and the stable point corresponds to the BHDE dominated epoch. For the stable point, since both $w_{D}$ and $q$ equal to $-1$, this point can behave as the cosmological constant. The attractor behavior and evolution curves of the stable point show that IBHDEHL model can describe the thermal history of the universe. Thus, IBHDEHL model is a viable cosmological model. In addition, for $1<\lambda<2$, the stable point can alleviate the coincidence problem since the ratio of the energy densities can approach to a constant.

In order to discriminate IBHDEHL model from the standard $\Lambda$CDM model, we apply the statefinder analysis method to IBHDEHL model. The statefinder diagrams show that the evolution curves of IBHDEHL model will eventually approach to the $\Lambda$CDM fixed point in $r-s$ phase plane and the de Sitter expansion fixed point in $r-q$ phase plane. Furthermore, the statefinder diagrams indicate that the interaction between the pressureless matter and BHDE has a significant effect on the evolution of the universe, and the IBHEDHL model can be distinguished from the standard $\Lambda$CDM model by this method.

Finally, we discuss the turning point of Hubble diagram in BHDE models. For BHDEF model, the turning point is nonexistent under the current observational results. For BHDEH model, this point vanishes for $\Delta>0.7$, and this point can also be avoided for $\Delta\sim 0.1$ in which the turning point occurs in the region $z\sim -1$ which means $a\rightarrow \infty$. When $\Delta \leq 0.02$, this point disappears. These results indicate that the turning point in Hubble diagram does not exist in BHDE models under the acceptable conditions.

\begin{acknowledgments}

This work was supported by the National Natural Science Foundation of China under Grants Nos. 11865018, 11865019, the Foundation of the Guizhou Provincial Education Department of China under Grants Nos. KY[2018]312, KY[2018]028, the Doctoral Foundation of Zunyi Normal University of China under Grants No. BS[2017]07.

\end{acknowledgments}

\end{document}